\documentclass[onecolumn,floatfix,showpacs,superscriptaddress]{revtex4-2}
\usepackage{mathtools}
\usepackage{bm}
\usepackage{dsfont,amsthm,amsbsy}
\usepackage{verbatim}
\usepackage{amssymb}
\usepackage{amsmath}
\usepackage{bbm}
\usepackage{graphicx}
\usepackage{epstopdf}
\usepackage{subfigure}
\usepackage{natbib}
\usepackage{epsfig}
\usepackage{amsfonts}
\usepackage{mathrsfs}
\usepackage{sidecap}
\usepackage{lipsum}
\usepackage[toc,page,title,titletoc,header]{appendix}
\usepackage[colorlinks,linkcolor=blue,citecolor=blue,anchorcolor=blue,urlcolor=blue]{hyperref}
\usepackage{hyperref}
\usepackage{resizegather}
\usepackage{tikz}
\usepackage{float}
\usepackage{mathbbol}
\usepackage[normalem]{ulem}
\usepackage{cancel}
\usepackage{upgreek}

\newcommand{\bl}{\begin{aligned}}
\newcommand{\el}{\end{aligned}}
\def\be{\begin{equation}}
\def\ee{\end{equation}}

\def\bi{\begin{itemize}}
\def\ei{\end{itemize}}
\def\bn{\begin{enumerate}}
\def\en{\end{enumerate}}
\def\bea{\begin{eqnarray}}
\def\eea{\end{eqnarray}}
\def\no{\nonumber}
\def\ba{\begin{array}}
\def\ea{\end{array}}
\def\bd{\begin{displaymath}}
\def\ed{\end{displaymath}}

\def\tr{{\rm tr}}

\begin{document}

\title{Dynamics of Steered Quantum Coherence and Magic Resource under Sudden Quench}

\author{Saeid Ansari}
\email{ansari@bzte.ac.ir}
\affiliation{Department of Engineering Sciences and Physics, Buein Zahra Technical University, Buein Zahra, 34518-66391, Iran}

\author{Alireza Akbari}
\affiliation{Institut für Theoretische Physik III, Ruhr-Universität Bochum, D-44801 Bochum, Germany}
\affiliation{Beijing Institute of Mathematical Sciences and Applications (BIMSA), Huairou District, Beijing, 101408, P. R. China}

\author{R. Jafari}
\affiliation{Department of Physics, Institute for Advanced Studies in Basic Sciences (IASBS), Zanjan, 45137-66731, Iran}
\affiliation{School of Nano Science, Institute for Research in Fundamental Sciences (IPM), Tehran, 19395-5531, Iran}
\affiliation{Department of Physics, University of Gothenburg, SE 412 96 Gothenburg, Sweden}
\begin{abstract}
  We explore the dynamics of $l_1$-norm of steered quantum coherence (SQC), steered quantum relative entropy (SQRE), and magic resource quantifier (QRM)  in the one-dimensional  XY  spin chain
in the presence of time dependent transverse magnetic field. We find that the system's response is highly sensitive to the initial state and magnetic field strength.
We show that the dynamics of SQC, SQRE and MRQ revealing the critical point associated with equilibrium quantum phase transition (QPT) of the system.
All quantities show maximum at QPT when the initial state is prepared in the ferromagnetic phase. Conversely, they undergo abrupt changes at quantum critical point if the initial state of the system is paramagnetic.
%
Moreover, our results confirm that, when quench is done to the quantum critical point, the first suppression (revival) time scales linearly with the system size, and remarkably, its scaling ratio remains consistent for all quenches, irrespective of the initial phase of the system.
These results highlight the interplay between the quantum information resources and dynamics of quantum systems away from the equilibrium. Such insights could be vital for quantum information processing and understanding non-equilibrium phenomena in quantum many-body systems.
\end{abstract}


\maketitle

\section{Introduction}
The study of quantum many-body systems driven out of equilibrium has become increasingly important from various
perspectives~\cite{Polkovnikov,Cazalilla,Heyl2018,Happola,Huang2006,Bayat:2010aa,metrology1,Bollinger,Bayat:2014aa,Raussendorf,Bayat:2007aa}.
 It plays a crucial role in investigating quantum information, particularly in the manipulation of coherence and entanglement dynamics. Understanding the notion of universality in non-equilibrium regimes, where the general principles of equilibrium systems do not apply, is also of great significance.
Advancements in experimental setups, such as ultra-cold atomic gases, optical lattices, and ion traps have provided 
 with the means to probe the non-equilibrium dynamical properties of closed many-body quantum systems in laboratories~\cite{Lamacraft,Gedik,Mandel,Bloch:2005aa,Treutlein,Cramer:2013aa,Leibfried}.
 The impact of equilibrium phase transitions on the modified state has allowed 
  to search for universal characteristics in non-equilibrium dynamics analogous to equilibrium phase transitions~\cite{Jafari2015,Jafari2016,Sharma2015,Montes2012,Sacramento2014,Jafari2017}.

Quantum quench, a method that involves suddenly changing external parameters of a closed quantum system, is a common approach to drive the system out of equilibrium~\cite{Greiner2002,Calabrese2006,Mitra:2018aa,NAG:2013aa}.
 This leads to unitary time evolution, and the resulting out-of-equilibrium quantum many-body models deviate from the established  principles of equilibrium systems.
The exploration of these post-quench dynamics and the identification of critical points have been tackled through several concepts, including the Loschmidt echo, Landau-Zener formalism, Kibble-Zurek mechanism, and dynamical quantum phase transitions~\cite{Gorin,Jacquod,Clark,Zurek,Heyl2013}.
Quantum coherence, arising from quantum state superposition, plays a fundamental role in quantum mechanics and has wide-ranging applications in quantum biology, thermodynamics, optics, and information
theory~\cite{Fazio2008,Eisert:2010aa,Plenio,Rebentrost,Lloyd,Huelga,Li2012,Lambert2013,Narasimhachar,Lostaglio1,Lostaglio2,Gour,Aberg,Cwiklinski,Ficek,Baumgratz,Girolami,Streltsov}.
It can be utilized to detect quantum phase transition (QPT) in many-body systems~\cite{Sachdev,Chakrabarti,Continentino,Osterloh2002,Karpat,Invernizzi,Sun}.
Approaches such as $l_1$-norm coherence and relative quantum coherence have been introduced for measuring quantum
coherence~\cite{Xi2015,Hu2016,Rana2016}, which serve as the foundation for various studies~\cite{chen,Qin2018,Gu2003}.
The utilisation of steering has extended the concepts of $l_1$-norm coherence and relative quantum coherence to incorporate long-range interactions in bipartite systems \cite{Mondal}.
 This  framework enables the exploration of  QPTs within quantum spin chains~\cite{Hu2020}. 
Additionally, a novel magic resource quantifier was recently proposed, based on the $l_1$-norm of characteristic functions of quantum states~\cite{Dai2022}.
Unlike some existing quantifiers, this new magic resource quantifier can be easily computed and is well-defined in all dimensions. Magic resources are essential for fault-tolerant quantum computation, and their
 straightforward calculation encourage researchers to explore its behaviour in quantum spin systems' critical points and its potential for detecting QPTs~\cite{Shor,Preskill1997,Knill,Bravyi2005,Campbell2017,Gottesman1998,Veitch2013,Bravyi2016,Howard2017,Ahmadi2018,Heinrich2019,Seddon2019,Bravyi2019,Wang2020,Seddon2021,Heimendahl2021,Liu2022}.
The role of this magic resource quantifier has been investigated in the context of the  XY  spin chain with three-spin interactions (XYT), and it has been demonstrated that the quantifier undergoes a sudden change around the critical points, thereby effectively signalling the presence of quantum critical points~\cite{Fu2022}.

In this paper, we focus on the one-dimensional XY-spin model in a transverse magnetic field.
We abruptly switch the external field to a new value at $t=0$ and investigate the time evolution of the system
using steered quantum coherence (SQC), steered quantum relative entropy (SQRE) and the dynamics of the magic resource quantifier (MRQ).

As previously mentioned, all of these quantifiers play significant roles in quantum information and quantum computation. Investigating the dynamics of these quantum resources holds promise for gaining valuable insights into understanding systems away from equilibrium.

The objective is to investigate whether the evolution of these quantities can function as a potential markers of QPTs within the system.
Furthermore, gaining an understanding of the universal behaviour of quantum information resources in non-equilibrium scenarios holds great potential for advancing our comprehension of quantum systems' descriptions.
By studying the non-equilibrium behaviour through SQC and MRQ and their relation to the equilibrium phase transition, we aim to detect the critical points.
This investigation could provide valuable insights into the role of steered quantum coherence and magic resource dynamics in detecting the QPTs and their potential applications in quantum information processing.

\section{Sudden quench in the transverse field XY model}
As a result of the quantum quench, the system experiences non-equilibrium dynamics since it starts from an initial state that is no longer an eigenstate of the new Hamiltonian.
The time evolution after the quench can lead to various interesting phenomena, such as the generation of quantum entanglement,
 the emergence of quantum phase transitions,
and the development of complex quantum correlations..
While they are essential to understanding the nature of quantum coherence, entanglement, and other non-equilibrium manifestations in quantum systems, they are also beneficial for understanding
the dynamics and  critical behaviours~\cite{Mishra2016,Jafari2017,Jafari:2010aa, Jafari:2020aa,Jafari:2019aa,Mishra:2018aa}.
Consequently, these phenomena are extensively investigated in both theoretical and experimental realms, aiming to decipher the behaviour of quantum many-body systems in scenarios when they deviate from equilibrium \cite{Xi2015,Hu2016,Rana2016,Qin2018,chen,Qin2018,Gu2003}.
The most  simplest paradigm is sudden quench, where a closed system is pushed out of equilibrium by a sudden change in the Hamiltonian which controls the time evolution of the system.
\\

Let's consider the general Hamiltonian as ${\cal H}(\lambda)$, where $\lambda$ is a tunable parameter. At an initial time $t<t_0$, the system is prepared in the ground state of the (pre-quenched) Hamiltonian ${\cal H}(\lambda=\lambda_i)$. Then, at time $t=t_0$, the parameter $\lambda$ is suddenly modified to a new value $\lambda_f$, causing the Hamiltonian to transform to ${\cal H}(\lambda=\lambda_f)$ (post-quenched Hamiltonian). As a consequence, the system evolves in time according to the post-quench Hamiltonian.
In the following, we choose the external magnetic field as the quench parameter in the XY model, by switching its value $h_i  \rightarrow h_f$ at $t=0$.
The Hamiltonian of the one-dimensional quantum spin  XY  model with $N$ sites and subject to the time-independent external transverse magnetic field
$h(t)=h_0+(h_1-h_0)\Theta(t)$, where $\Theta(t)$ is the Heaviside step function (which means the magnetic filed suddenly changes from $h_0$ to $h_1$ at $t=0$), can be written as follows:
\begin{equation}
\label{hamiltonian}
{\cal H}
=
-\frac{J}{2} \sum_{i=1}^{N}
[
(1+\gamma)\sigma_{i}^{x} \sigma_{i+1}^{x}+(1-\gamma)\sigma_{i}^{y} \sigma_{i+1}^{y}
]
- h(t) \sum_{i=1}^{N}\sigma_{i}^{z}.
\end{equation}
Here $J$ denotes the strengths of time-independent exchange coupling between the nearest-neighbor spins in the chain with anisotropy $\gamma$.
Moreover, $h$ is the strength of magnetic field, and $\sigma_i^{\alpha=x,y,z}$ are the Pauli matrices, representing the spin operators at site $i$,
and $\sigma_i^0$ is $2\times 2$ unit matrix.

Indeed, the Jordan-Wigner transformation
is a powerful technique in quantum many-body physics, particularly for one-dimensional quantum spin chains. It allows us to map the spin operators to fermionic operators, leading to a quadratic Hamiltonian that can be diagonalized exactly in momentum space.
In the time-independent magnetic field, the ground state of the model characterized by a QPT 
that takes place at the critical point $h_{c}=J$  \cite{McCoy:1968aa,phase}.
The order parameter is the magnetization $\langle M^{x,y}\rangle$ which differs from zero for $h<h_{c}$ and zero otherwise.
The ground state of the system is paramagnetic when $h\rightarrow \infty$, where the spins aligned in $z$ (the magnetic field) direction.
For the other extreme case when $h\rightarrow0$,
 the ground state of the system is ferromagnetic and the spins are all aligned
in the $x$ ($y$) direction when $\gamma>0$ ($\gamma<0$).
This means that in both cases the state is minimally entangled. QPT takes place
at zero temperature as the thermal fluctuations destroy the quantum correlations in the ground state of the system.

If the system prepared initially in its ground state, after the sudden quench, the symmetrical two-spin reduced density matrix $\rho_{l,m}$ for two spins located at nearest-neighbor sites $l$ and $m$ in the chain will exhibit a specific structure known as the X-form~\cite{Osterloh2002,Sadiek2010}, which  expressed as follows
\begin{equation}
\rho_{l,m}=\frac{1}{4}
\left(
\begin{array}{cccc}
 \rho_{11} &0 &0 & \rho_{14} \\
 0 & \rho_{22} & \rho_{23} &0 \\
 0 & \rho_{23}^{\ast} & \rho_{33} &0 \\
  \rho_{14}^{\ast} &0 &0 &\rho_{44} \\
\end{array}
\right)
.
\label{eq:rhot}\end{equation}
It has six distinguished nonzero elements
that can be addressed in terms of the magnetization and two-point correlation functions. They are given as 
%
\begin{eqnarray}
\bl
\nonumber
 \rho_{11}=&1+ 2\big\langle M^z_{l}\big\rangle+\big\langle \sigma^{z}_{l} \sigma^{z}_{m} \big\rangle ,
\quad
 \rho_{22}=\rho_{33}=1-\big\langle \sigma^{z}_{l} \sigma^{z}_{m} \big\rangle ,
\\
 \rho_{44}=
&
1- 2\big\langle M^z_{l}\big\rangle+\big\langle \sigma^{z}_{l} \sigma^{z}_{m} \big\rangle ,
\quad
 \rho_{23}=
\big\langle \sigma^{x}_{l} \sigma^{x}_{m} \big\rangle+\big\langle \sigma^{y}_{l} \sigma^{y}_{m} \big\rangle,\\
\rho_{14}=&\big\langle \sigma^{x}_{l} \sigma^{x}_{m} \big\rangle-\big\langle \sigma^{y}_{l} \sigma^{y}_{m} \big\rangle ,
\el
\end{eqnarray}
where $M^z$ is the system transverse magnetization in the $z$-direction, defined as average value of all spins' magnetization
\bea
\bl
M^{z}=\frac{1}{N}\sum_{j=1}^{N} \sigma_{j}^{z}.
\el
\eea
and the expectation value of the average values can be expressed in terms of the density matrix $\rho$. For a given operator $\hat{\cal O}$, the expectation value $\langle \hat{\cal O} \rangle$ is defined by
$\langle \hat{\cal O} \rangle=\frac{Tr[\rho \hat{\cal O}]}{Tr[\rho]}.$
%
%
%
Employing the Wick Theorem at zero temperature,
enables us to derive the nearest-neighbour spin-spin correlation functions as follows:~\cite{Barouch1,Barouch2,Sadiek2010,Huang2006,Mishra2016}

\bea
\bl
\label{eqApp4}
&
\langle S^{x}_{l} S^{x}_{l+1} \rangle
 =
 \frac{1}{4}F_{l,l+1} ,
\quad
\langle S^{y}_{l} S^{y}_{l+1} \rangle
=
\frac{1}{4}F_{l+1,l} ,
\\
&
\langle S^{z}_{l} S^{z}_{l+1} \rangle =
\frac{1}{4}
\Big\{
F_{l, l}\times F_{l+1,l+1}
-Q_{l,l+1}\times G_{l,l+1}
-F_{l+1,l}\times F_{l,l+1}
\Big\}.
\el
\eea
%
In which by considering  $\phi_p = 2\pi p/N$, $\delta_p = 2\gamma \sin (\phi_p)$,  $\delta_h=h_{0}- h_{1}$,  
and   setting
\begin{equation}
\label{eqApp6}
\Gamma_h[h(t)]=
\Big\{[J\cos \phi_{p} + h(t)]^{2}+\gamma^2 J^2 \sin^2\phi_{p}\Big\}^{\frac{1}{2}},
%
\end{equation}
we define
%
%
\bea
\no
\label{eqApp5}
Q_{l, m}
=
-G^{*}_{l, m}
&=
&
\frac{1}{N} \sum_{p=1}^{N/2} \cos[(m-l)\phi_{p}]
\times
 \biggl\{ 2+ iJ \delta_h \delta_{p}\tan[(m-l)\phi_{p}]\frac{\sin[4t\Gamma_h(h_{1})]}{\Gamma_h(h_{1})\Gamma_h(h_{0})}\biggr\} ,
\\
\eea
and
\bea
\no
F_{l, m}
&
=
&
\frac{1}{N}
 \sum_{p=1}^{N/2}
\frac{1}{\Gamma_h^2(h_{1})\Gamma_h(h_{0})}
\cos[(m-l)\phi_{p}]
\times
\\
&&
\Biggl\{
\biggl[
J ^2
\delta_h
\delta^2_{p} \sin^2[2t\Gamma_h(h_{1})]
+ 2\Gamma_h^2(h_{1})
(J\cos\phi_{p}+h_{0})
\biggr]
\\\no
&&
+
 \delta_{p} J
\tan[(m-l)\phi_{p}]
\times
\biggl[
\Gamma_h^2(h_{1})
+
2
\delta_h
(J\cos\phi_{p}+h_{1})
\sin^2[2t\Gamma_h(h_{1})
]
\biggr]
\Biggr\}.
\eea
%


%
\section{Steered quantum coherence}\label{sec:2}
Quantum coherence is a fundamental concept in quantum mechanics, closely related to the superposition principle of quantum pure states. It forms the basis of many quantum phenomena.
Steering, initially proposed by Schr{\"o}dinger~\cite{Schodinger} as a solution to the Einstein-Podolsky-Rosen (EPR) paradox~\cite{EPR}, represents a significant nonlocal quantum correlation. In the context of steering, when two subsystems are entangled, performing local measurements on one subsystem can influence the quantum state of the other~\cite{Wiseman2007}. The ability of this unique to remotely control the state of a distant system is achievable only through quantum correlation. A distinguishing feature of steered quantum coherence is its applicability in long-range spin-spin interactions, setting it apart from other approaches such as entanglement and quantum discord.

Due to this advantage, SQC has been the subject of both theoretical investigations~\cite{Hu2016,Costa2016} and experimental interest~\cite{SQC-Lab1,SQC-Lab2,SQC-Lab3} in recent years. Notably, Hu {\it et al.}~\cite{Hu2020} successfully utilised SQC to explore the critical behaviour of quantum phase transitions in various one-dimensional quantum spin models, demonstrating its accuracy in determining the QPT points.
In addition,  
 SQC can  be used to study the time evolution of quantum systems driven out of equilibrium.
To address this question, we apply the sudden quantum quench method in the transverse field  XY  model.
Following the Ref.~\cite{Hu2020}, the $l_1$-norm steered quantum coherence, denoted as $C_{l_1}$, and the relative steered quantum coherence, denoted as $C_{\text{re}}$, can be expressed analytically in terms of the spin-spin density operator $\rho_{l,m}$, i.e.,
\bea
 \label{}
\bl
  C_{l_1}(\rho_{l,m})=
  &
     \langle M^z_{l}\rangle +\frac{1}{2}
   \Big(
   \langle \sigma^{x}_{l} \sigma^{x}_{m} \rangle
+\langle \sigma^{y}_{l} \sigma^{y}_{m}\rangle
   \Big)
   \\
&
   +
\frac{1}{2}
\Big(
\sqrt{\langle M^z_{l}\rangle^2+\langle \sigma^{x}_{l} \sigma^{x}_{m} \rangle^2}
            +\sqrt{\langle M^z_{l}\rangle^2+\langle \sigma^{y}_{l} \sigma^{y}_{m}\rangle^2}
            \Big),
  \\
   C_{\text{re}}(\rho_{l,m})=
   &
    2-H_2(C_1)-H_2(C_2) + H_2
   \Big(
   \frac{1+\langle M^z_{l}\rangle}{2}
   \Big)
   \\
       &
-\frac{1}{2} \Big(1+\langle M^z_{l}\rangle\Big)H_2(C_3) -\frac{1}{2}\Big(1-\langle M^z_{l}\rangle\Big)H_2(C_4).
\nonumber
 \el
\eea
Here $H_2(x)$ is the Shannon entropy function~\cite{Roy2020}, defined by
\begin{equation}\label{}
H_2(x)=-x\log x-(1-x)\log (1-x),
\end{equation}
and $C_i$ ($i=1,2,3,4$) are given as follows
\begin{equation} \label{eq2b-11}
 \begin{aligned}
  & C_1=\frac{1}{2}\Big(1 + \sqrt{\langle M^z_{l}\rangle^2+\langle \sigma^{x}_{l} \sigma^{x}_{m} \rangle^2}\Big) ;
  \quad
  C_2=\frac{1}{2}\Big(1 + \sqrt{\langle M^z_{l}\rangle^2+\langle \sigma^{y}_{l} \sigma^{y}_{m} \rangle^2}\Big);
    \\
  & C_3=\frac{1}{2} \Big(1+\frac{|\langle M^z_{l}\rangle+\langle \sigma^{z}_{l} \sigma^{z}_{m} \rangle|}{1+\langle M^z_{l}\rangle}\Big) ;
  \quad\quad\;
   C_4=\frac{1}{2} \Big( 1+\frac{ |\langle M^z_{l}\rangle+\langle \sigma^{z}_{l} \sigma^{z}_{m} \rangle|}{1-\langle M^z_{l}\rangle} \Big).
 \end{aligned}
\end{equation}
Note that $C_{\text{re}}$ offers a normalized measure of steered coherence by dividing the sum of absolute off-diagonal elements by the maximum such element and then subtracting 1.

%
\section{Magic resource quantifier}\label{sec:3}
The advent of quantum computers has sparked significant research efforts, driven by their ability to perform computations at much faster speeds than classical computers. However, achieving fault-tolerant universal quantum computation requires additional resources, including the elusive ``magic resource"~\cite{Bravyi2005}. Consequently, quantifying this magic resource has  
been leading to the evolution of various magic resource quantifiers~\cite{Gottesman1998,Veitch2013,Veitch2013,Bravyi2016,Howard2017,Ahmadi2018,Heinrich2019,Seddon2019,Bravyi2019,Wang2020,Seddon2021,Heimendahl2021,Liu2022}.
Recently, a novel magic resource quantifier was proposed by Dai {\it et al.}~\cite{Dai2022}, based on the characteristic functions of quantum states. This magic resource quantifier is both easy to compute and well-defined in all dimensions, making it an attractive tool for quantifying the magic resource. Encouraged by these developments, Fu and collaborators~\cite{Fu2022} suggested employing this new magic resource quantifier to explore critical points in many-body quantum systems. Their work demonstrated that the presence of the magic resource plays a crucial role in signaling the critical points of quantum spin systems.
In the context, our attention is directed towards a specific magical quantifier:
\begin{equation}
\label{eq8}
M(\rho)=\sum_{k} |c_{\rho}(k,l)|,\:\: k,l \in \mathbb{Z}_d,
\end{equation}
which corresponds to the $l_1$-norm of the characteristic function $c_\rho(k, l)$ for a given quantum state $\rho$.
Through analysing it, it becomes possible to evaluate the quantum magic resource necessary for fault-tolerant universal quantum computation and various other quantum tasks.
The density operator  for $n$-qubit state
can be obtained from
\begin{equation}\label{}
M(\rho)=\sum_{k} |\tr(\rho\sigma_{\textbf{k}})|
\end{equation}
with $\sigma_{\textbf{k}}=\sigma_{k_1} \otimes \cdots \otimes \sigma_{k_j} \otimes\cdots \otimes \sigma_{k_n}$, $k_j=0,x,y,z$.
%
For the nearest neighbor reduced two-spin state $\rho_{i,i+1}$ of the ground state, relation for the $l_1$-norm of the characteristic function
simplifies as follows
\begin{equation}\label{}
M(\rho_{i,i+1})=\sum_{s,t}{ |c_{\rho_{i,i+1}}(s,t)|}=\sum_{s,t}{ |\tr(\rho_{i,i+1}\sigma_i^s\otimes\sigma_{i+1}^t)|},
\end{equation}
where setting $s,t\in\{0,x,y,z\}$ will give a fourth-order square matrix for $c_{\rho_{i,i+1}}(s,t)$ as
\begin{equation}
c_{\rho_{i,i+1}}(s,t)=
\left(
\begin{array}{cccc}
 1 &0 &0 & \langle \sigma^{z}\rangle \\
 0 & \langle \sigma^{x}_{i} \sigma^{x}_{i+1} \rangle & 0 &0 \\
 0 & 0 & \langle \sigma^{y}_{i} \sigma^{y}_{i+1} \rangle &0 \\
  \langle \sigma^{z}\rangle &0 &0 &\langle \sigma^{z}_{i} \sigma^{z}_{i+1} \rangle \\
\end{array}
\right)
,
\label{eq:rhot}\end{equation}
that leads to following expression for magic resource quantifier~\cite{Fu2022}
\begin{equation}\label{}
\bl
M(\rho_{i,i+1})=1+\vert \langle \sigma^{x}_{i} \sigma^{x}_{i+1} \rangle \vert+\vert \langle \sigma^{y}_{i} \sigma^{y}_{i+1} \rangle \vert+\vert \langle \sigma^{z}_{i} \sigma^{z}_{i+1} \rangle \vert+2\vert \langle \sigma^{z}\rangle \vert
.
\el
\end{equation}
%

%
\section{Numerical Results and discussions}
%
The investigation of various quench scenarios and system sizes contributes to a comprehensive understanding of the dynamics and properties of  steered quantum coherence and magic resource, 
can shed light on the general behavior of quantum systems undergoing sudden changes in external parameters.
In this section our objective is to examine whether the relative entropy and $l_1$-norm dynamics and the evolution of the magic resource quantifier ,subsequent to a sudden quench in the transverse magnetic field, can serve as an indicators of the QCP within the model. Furthermore, we aim to explore the correlations between these measures and the QPTs that occur.
\\

%
\begin{figure}[t]
\centerline{
\includegraphics[width=\linewidth]{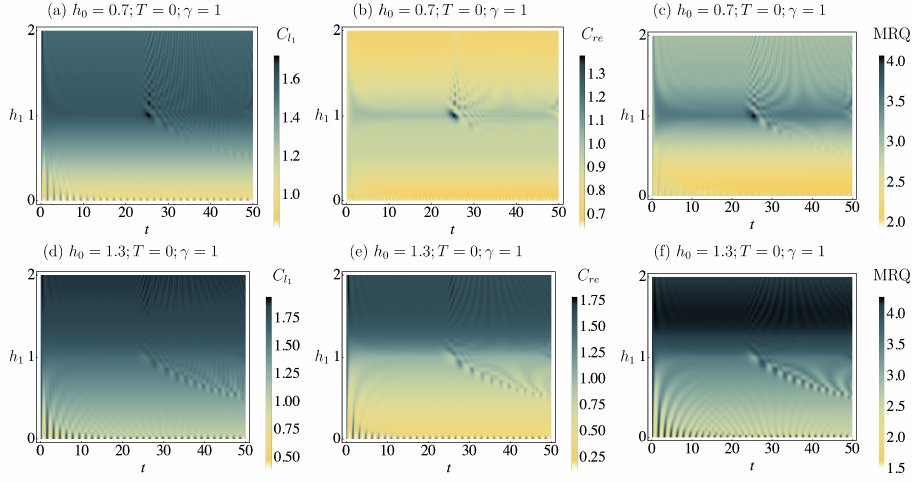}
}
\caption{
Depict the density plots of the dynamics of quantum measures versus $t$ and $h_1$
in the Ising model ($\gamma=1$) at zero temperature.
 Panels (a) and (d):  $l_1$-norm of coherence ($C_{l_1}$),
 panels (b) and (e):  relative entropy ($C_{\text{re}}$),
and
panels (c) and (f): magic resource quantifier (MRQ).
By considering a quench from $h_0=0.7<h_c$  in (a-c) panels,
 and  from $h_0=1.3 > h_c$
 in  (d-f) panels.
}
\label{Fig1}
\end{figure}
%

%
\begin{figure}[t]
\centerline{
\includegraphics[width=\linewidth]{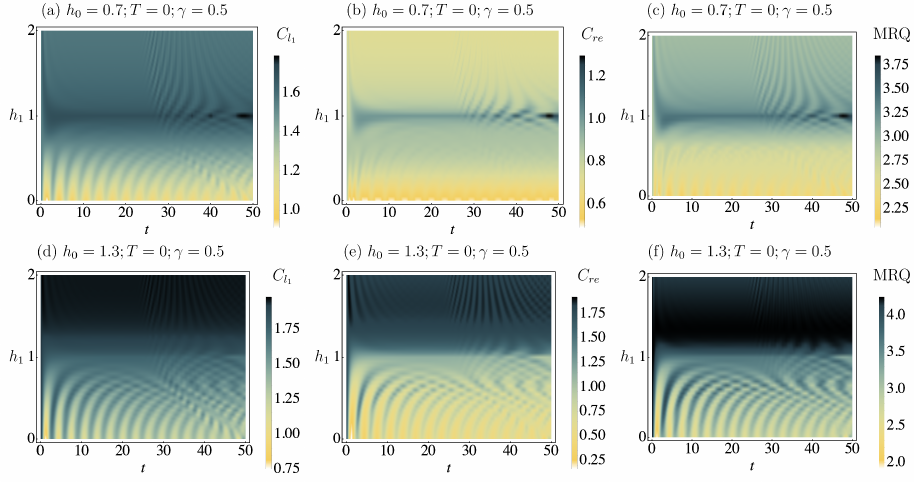}

}

\caption{
 Density plots of the dynamics of quantum measures for anisotropic system with $\gamma=0.5$ at zero temperature.
The plots order arranges the same as Fig. \ref{Fig1},  while
considering a quench from $h_0=0.7<h_c$ (a-c panels) and from $h_0=1.3>h_c$ (d-f panels).
}
\label{Fig2}
\end{figure}
%

We begin by performing quenches from an initial magnetic field $h_0$ to any $h_1$, regardless of whether $h_0<h_c$ or $h_0>h_c$.
This initial exploration allows us to observe the dynamics of $l_1$-norm of SQC, relative entropy and the magic resource quantifier for different quench scenarios.
Subsequently, we focus on quenching the system to the critical point $h_c$ and study the dynamics of anisotropic spin systems with various system sizes.
%

%
We first consider the transverse field Ising model ($\gamma=1)$ .
Fig.~\ref{Fig1} illustrates the dynamics of relative entropy, $l_1$-norm coherence and magic resource quantifier for chains with lengths $N=100$. The panels in the first row, 
(a), (b) and (c),
show dynamics of $l_1$-norm steered quantum coherence ($C_{l_1}$),
relative entropy ($C_{\text{re}}$) and magic resource quantifier (MRQ) for the case that the initial magnetic field is below the transition field ($h_0=0.7<h_c=1$).
The observations reveal a distinct pattern: for zero $h_{1}$, where the spins are fully aligned along in the $x$ direction, all considered quantities display an oscillatory pattern over time.
Upon introducing a magnetic field, the magnitudes of these measures grow with increasing $h_{1}$ until they peak at the equilibrium critical point $h_{1}=h_{c}=1$.
By surpassing $h_{c}$, the magnitudes of $C_{\text{re}}$ and MRQ gradually diminish with the influence of the magnetic field, while the reduction in $C_{l_1}$ beyond $h_c$ is marginal, rapidly converging to a saturated value. This pattern indicates that when the system is initially prepared in a ferromagnetic phase ($M^{x}\neq0$), the maximum values of all these quantities coincide with the equilibrium critical point, which serves as an accurate representation of the system's critical point.

To provide a more comprehensive insight into dynamics of the aforementioned quantities beyond the transition field, the second row of Fig. \ref{Fig1} depicts the variations of the quantum measures
by assuming  $h_0=1.3$, where the system's initial state is situated in the paramagnetic phase ($M^{x}=0$).
 Evidently, for $h_1>h_c$, all measured quantities reach saturation and display a distinct decline in the vicinity of the critical point $h_c=1$. These results substantiate the notion that dynamical steered quantum coherence has the capability to discern the critical point of system.

To complete picture, we also present density plots of the mentioned quantum measures for the anisotropic case $\gamma=0.5$ in Figure \ref{Fig2}. 
 As anticipated, when the system's initial state is configured in the ferromagnetic case ($h_0=0.7$), all quantities attain their maxima precisely at the system's critical point. Furthermore, when the system commences in the paramagnetic phase ($h_0=1.3$), as expected, the quantities undergo a steep reduction at the critical point, thus confirming the consistency of this behaviour across different initial conditions.

%

\subsection{Quantum Quench to the Critical Point}
\begin{figure}[t]
\centerline{
 \includegraphics[width=\linewidth]{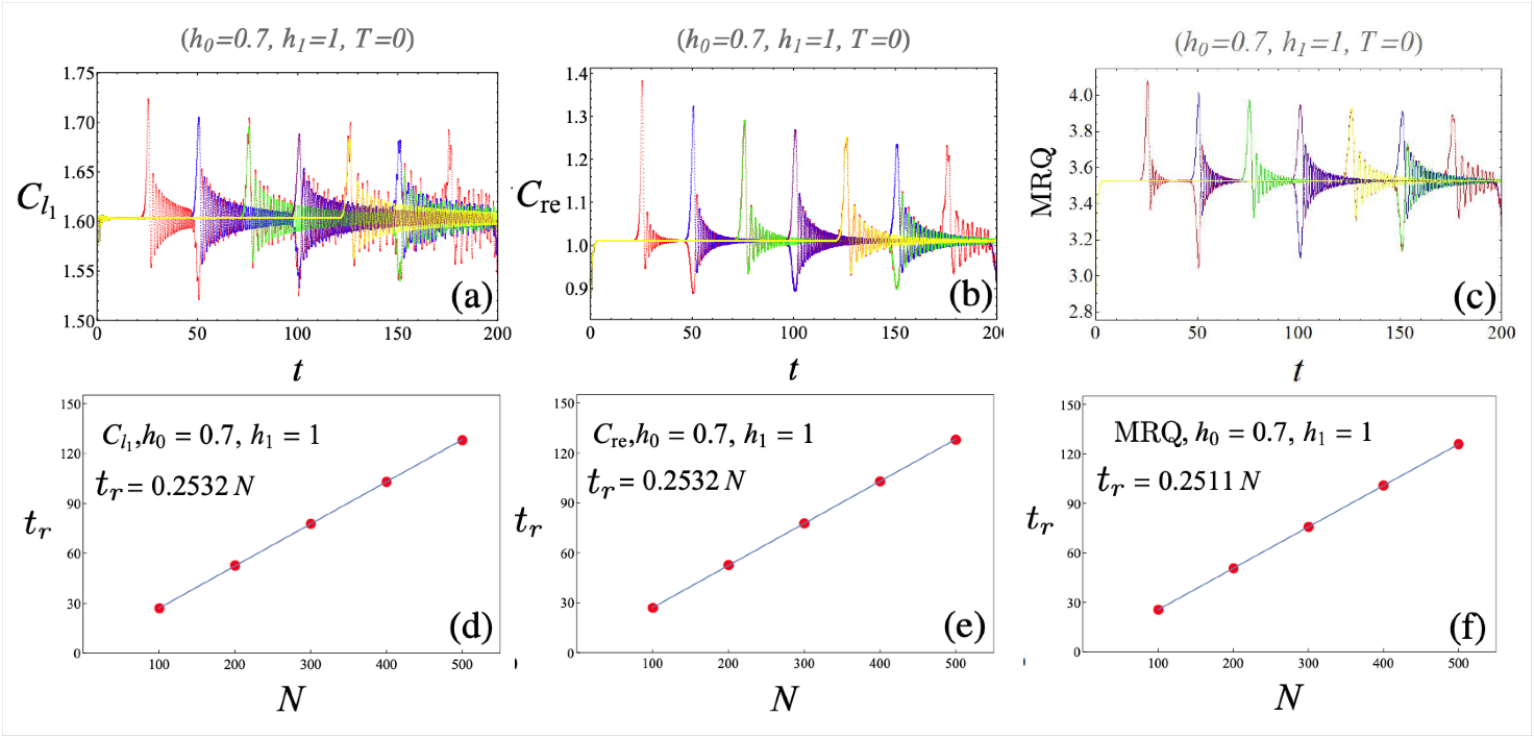}
 }
    \caption{
The time evolution of quantum measures is illustrated for a quench to the critical point from below ($h_0=0.7\rightarrow h_1=h_c=1$)   at zero temperature for different system size:
 (a) the evolution of $l_1$-norm of quantum coherence, (b) the relative entropy, and (c) the magic resource quantifier.
We consider system sizes of $N=100, 200, 300, 400$, and $500$, each represented by distinct coloured lines. (d), (e) and (f) show the linear behaviour of the first suppression-time (revival-time), $t_r(N)$, versus the system size for $C_{l_1}$, $C_{\text{re}}$ and MRQ, respectively.
}
\label{Fig4}
\end{figure}

\begin{figure}[t]
\centerline{
 \includegraphics[width=\linewidth]{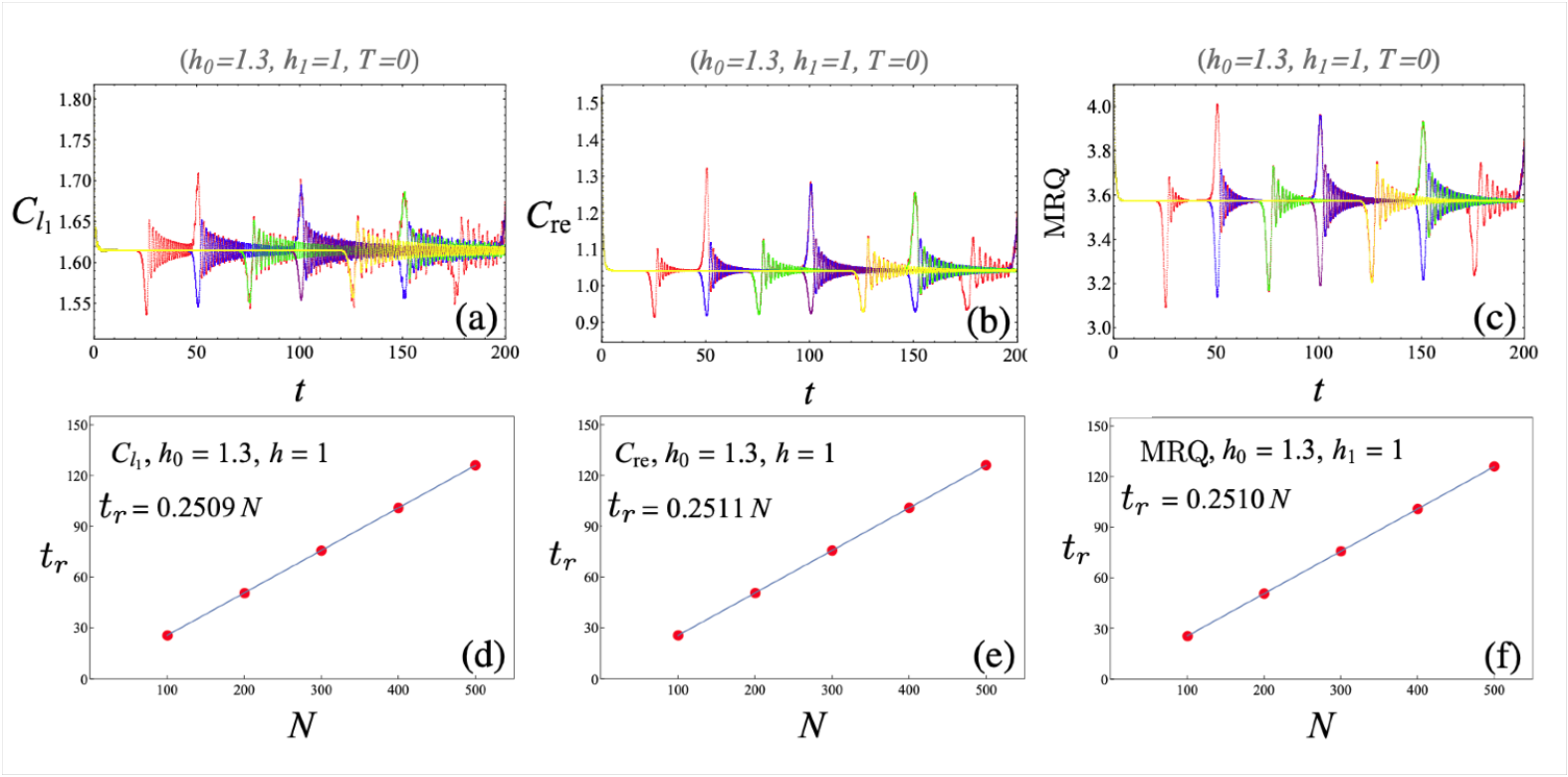}
    }
      \caption{Same plots as Fig.~\ref{Fig4}, but for a quench to the critical point from above ($h_0=1.3\rightarrow h_1=h_c$).}
\label{Fig5}
\end{figure}

%
In our next analysis, we focus on quenching the system to the critical point ($h_1 = h_c$). We first set $h_0=0.7$ (where $h_0<h_c$) and present the time evolution of $C_{l_1}$, $C_{\text{re}}$, and 
MRQ in Fig.~\ref{Fig4}. We then consider $h_0=1.3$ (where $h_0>h_c$) in Fig.~\ref{Fig5}.
 %
%

Here, to ensure a comprehensive analysis,  we delve into  the temporal evolution of these quantities across systems of varying sizes. Specifically, we consider system lengths of $N=100, 200, 300, 400$, and $500$, each represented by distinct coloured lines in Fig.~\ref{Fig4}(a)-(c) and Fig.~\ref{Fig5}(a)-(c).
As is clear, in a very short time all quantities change rapidly from the equilibrium state to
their average value which they oscillate around. Moreover, all quantities show suppressions and revivals as
deviations from the average value.
Our aim is to investigate how the system size affects the first revival/suppression time, $t_r$, of these quantum measures.
In this respect,  in Fig.~\ref{Fig4}(d)-(f) and Fig.~\ref{Fig5}(d)-(f), we show that the revival/suppression time, $t_r$, is linearly proportional to the system size, $N$. This relationship can be succinctly expressed as follows:
\begin{equation}
t_r(N) = \tau_r N,
\end{equation}
where $\tau_r$ is the scaling ratio. For $l_1$-norm coherence ($C_{l_1}$) and relative steered quantum coherence ($C_{\text{re}}$), we find $\tau_r = 0.2532$, while for the magic resource quantifier (MRQ) we obtain $\tau_r = 0.2511$.
 We have also explored the quantities when the system is initially prepared at the critical point $h_0=h_c$.
We have noted that the dynamics of these quantities for a quench from the critical point, i.e., $h_0=h_c$, exhibit qualitative similarity to the dynamics observed when $h_0=h_c$.

Our calculations show that $t_r$ and $\tau_r$ are the same for all quenches to/from critical point and do not
depend on the phase of system where the system is prepared. This is the promised universality of $t_r$  which
shows that the size of quench and the phase of system, are ineffective.

%
%
%
\section{Conclusion}
%
The quantum measures, including the relative entropy of coherence, $l_1$-norm of coherence and magic resource quantifier,
serve as sensitive probes to identify quantum correlations  and
 reveal intriguing behaviours linked to  quantum phase transitions.
The novel magic resource quantifier holds significant promise in advancing our comprehension of quantum phase transitions and critical behaviours within complex quantum systems. Its prospective utilization stands to yield revolutionary breakthroughs in quantum technology, with far-reaching impacts on computation, communication, and information processing. Moreover, its quantification could  optimize the efficiency of  quantum algorithms, with potential applications spanning cryptography, material science, and drug discovery.
In essence, the emerging magic resource quantifier bears immense potential for shaping the trajectory of quantum technology and computation in the years to come.
%

%
We have analyzed the dynamical behaviour of $l_1$-norm of steered quantum coherence, steered quantum relative entropy, and magic resource quantifier in a 1D XY model that experiences a sudden quench in the transverse magnetic field.
Remarkably, we observe suppressions and revivals of quantum measures during the quenching process, particularly  in proximity to  the critical point.
We confirm  that the first suppression (revival) time scales linearly with the  system's size, regardless of quench parameters or the initial phase,
 highlighting the universality of quantum correlations in out-of-equilibrium many-body systems.
The observed universality  emphasizes the fundamental nature of quantum correlations and their relevance in quantum phase transitions.
Our work contributes substantially to the understanding of quantum systems experiencing sudden quenches, offering a comprehensive depiction of coherence and magic resource dynamics within such models.

In conclusion, our results demonstrate the significance of steered quantum coherence and the magic resource quantifier in predicting critical points and understanding quantum phase transitions. The robustness of our findings across different quench scenarios and anisotropic systems confirms the reliability and applicability of these measures in studying quantum correlations and their dynamics in a wide range of quantum spin models.

\section*{Acknowledgments}
A.~A. acknowledges the financial support from the German Research Foundation within the bilateral NSFC-DFG Project No. ER 463/14-1.

\appendix

\section{Basic Concepts of Steered Quantum Coherence}\label{secA1}
In this appendix, we give general analytical solution of the SQC. We consider a two-qubit state $\rho_{AB}$ shared by the subsystems A and B. The SQC is defined by local measurements of subsystem A and classical communication between A and B.
Suppose A performs one of the measurements $\{\sigma^\mu\}_{\mu=x,y,z}$ (Pauli operators) with an outcome $a\in\{0,1\}$ and communicates the choice with B.
Consequently, the state of qubit B undergoes a collapse (steering) to the ensemble states
$\{p_{\mu,a},\rho_{B|\Pi_\mu^a}\}$, where $p_{\mu,a}={\text{Tr}}(\Pi_\mu^a \rho_{AB})$
 denotes the probability of obtaining outcome $a$.
 Moreover,
   $\rho_{B|\Pi_\mu^a}={\text{Tr}_A}(\Pi_\mu^a \rho_{AB})/p_{\mu,a}$
   represents the conditional state of subsystem B,
   and $\Pi_\mu^a=[I_2+(-1)^a \sigma^{\mu}]/2$
stands for the measurement operator, with $I_2$ denoting the identity operator.

%
Once A completes all possible measurements with equal probability and randomly communicates the chosen observable $\sigma^{i}$ to subsystem B, B can then measure the steered coherence on its side. If subsystem B randomly selects one of the two remaining eigenbases of ${\sigma^j}; \; {j\ne i}$, the SQC is defined as the following averaged quantum coherence
\begin{equation}
 C^{na}(\rho(AB))=\frac{1}{2} \sum_{\mu\ne\nu,a} p_{\mu,a}C^{\sigma^{\nu}}(\rho_{B|\Pi_{\mu}^a}),
\end{equation}
which  represents the coherence of the steered state with respect to the eigenbases of $\sigma^\nu$.
Here we use $l_1$-norm of steered quantum coherence ($C_{l_1}^{\sigma^\nu}$) and the relative entropy of steered quantum coherence ($C_{(\text{re}}^{\sigma^\nu}$). By denoting $\{|\psi_i\rangle\}$ the eigenbases of $\sigma^\nu$, the analytical solutions can be written as \cite{Hu2020}
\begin{eqnarray}
\bl
 C_{l_1}^{\sigma^\nu}
 =
  \sum_{i\ne j}|\langle\psi_i|\rho|\psi_j\rangle|;
  \quad
  C_{\text{re}}^{\sigma^\nu}
  =
   \sum_{i}\langle\psi_i|\rho|\psi_i\rangle \log_2\langle\psi_i|\rho|\psi_i\rangle-S(\rho),
  \el
\end{eqnarray}
where $S(\rho)=-\text{Tr}\rho\log_2\rho$ is the von Neumann entropy.
Based on these formulas, one can compute the corresponding steered quantum coherence.

To calculate the SQC, we need to obtain the density operator $\rho_{i,i+r}$ for the spin pair ($i,i+r$), representing two spins with a distance $r$ in units of the chain constant.
In the Bloch representation, the density operator can be decomposed as:
\begin{equation}\label{}
\rho_{i,i+r}=\frac{1}{4} \sum_{\mu,\nu} t_{\mu,\nu} \sigma_i^\mu \otimes \sigma_{i+r}^\nu,  \quad  (\mu,\nu\in{0,x,y,z})
\end{equation}
 where $t_{\mu,\nu}=\text{Tr}(\rho_{i,i+r} \sigma_i^\mu \otimes \sigma_{i+r}^\nu)$.
Non-zero $t_{\mu,\nu}$ can be obtained in terms of the magnetization and the spin-spin correlation function as
\begin{equation}\label{}
t_{z0}=t_{0z}=\langle \sigma_z\rangle, \quad t_{\mu\mu}=\langle \sigma_i^\mu \sigma_{i+r}^\mu\rangle, \quad  (\mu,\nu\in{x,y,z}).
\end{equation}
Thus, the SQC can be determined using the expressions derived earlier and the non-zero values for the two-spin density operator
\begin{eqnarray}
\bl
  C_{l_1}^{na} (\rho_{i,i+r})
  =&
   \langle\sigma^z\rangle+\frac{1}{2}\left( \langle\sigma^x_i \sigma^x_{i+r}\rangle+  \langle\sigma^y_i \sigma^y_{i+r}\rangle+\sqrt{\langle\sigma^x_{i} \sigma^x_{i+r}\rangle^2+\langle\sigma^y_i \sigma^y_{i+r}\rangle^2} \right);
    \\
   C_{\text{re}}^{na} (\rho_{i,i+r})
   = &
    2-H_2(t_1)-H_2(t_2)
    -\Big[\frac{1+\langle\sigma^z\rangle}{2}\Big] H_2(t_3)
    -\Big[\frac{1- \langle\sigma^z\rangle}{2}\Big] H_2(t_4)
    \\ &
    +    H_2\left(\frac{1+\langle\sigma^z\rangle}{2}\right),
    \no
   \el
   \\
\end{eqnarray}
where $H_2(\cdots)$ denotes the binary Shannon entropy function and the $t_i$'s are given as
\begin{eqnarray}
\bl
  t_1
  &=
   \frac{1}{2}\left(1+\sqrt{\langle\sigma^z\rangle^2+ \langle\sigma^x_i \sigma^x_{i+r}\rangle^2} \right);
    \quad
    t_2 = \frac{1}{2}\left(1+\sqrt{\langle\sigma^z\rangle^2+ \langle\sigma^y_i \sigma^y_{i+r}\rangle^2} \right) ;
   \\
  t_3
  &= \frac{1+|\langle\sigma^z\rangle+ \langle\sigma^z_i \sigma^z_{i+r}\rangle|}{2[1+\langle\sigma^z\rangle]};
   \quad\quad\quad\quad\;\;
    t_4 = \frac{1+|\langle\sigma^z\rangle- \langle\sigma^z_i \sigma^z_{i+r}\rangle|}{2[1-\langle\sigma^z\rangle]}
    .
 \el
\end{eqnarray}


\bibliography{SQC-References}

\begin{thebibliography}{103}%
\makeatletter
\providecommand \@ifxundefined [1]{%
 \@ifx{#1\undefined}
}%
\providecommand \@ifnum [1]{%
 \ifnum #1\expandafter \@firstoftwo
 \else \expandafter \@secondoftwo
 \fi
}%
\providecommand \@ifx [1]{%
 \ifx #1\expandafter \@firstoftwo
 \else \expandafter \@secondoftwo
 \fi
}%
\providecommand \natexlab [1]{#1}%
\providecommand \enquote  [1]{``#1''}%
\providecommand \bibnamefont  [1]{#1}%
\providecommand \bibfnamefont [1]{#1}%
\providecommand \citenamefont [1]{#1}%
\providecommand \href@noop [0]{\@secondoftwo}%
\providecommand \href [0]{\begingroup \@sanitize@url \@href}%
\providecommand \@href[1]{\@@startlink{#1}\@@href}%
\providecommand \@@href[1]{\endgroup#1\@@endlink}%
\providecommand \@sanitize@url [0]{\catcode `\\12\catcode `\$12\catcode
  `\&12\catcode `\#12\catcode `\^12\catcode `\_12\catcode `\%12\relax}%
\providecommand \@@startlink[1]{}%
\providecommand \@@endlink[0]{}%
\providecommand \url  [0]{\begingroup\@sanitize@url \@url }%
\providecommand \@url [1]{\endgroup\@href {#1}{\urlprefix }}%
\providecommand \urlprefix  [0]{URL }%
\providecommand \Eprint [0]{\href }%
\providecommand \doibase [0]{https://doi.org/}%
\providecommand \selectlanguage [0]{\@gobble}%
\providecommand \bibinfo  [0]{\@secondoftwo}%
\providecommand \bibfield  [0]{\@secondoftwo}%
\providecommand \translation [1]{[#1]}%
\providecommand \BibitemOpen [0]{}%
\providecommand \bibitemStop [0]{}%
\providecommand \bibitemNoStop [0]{.\EOS\space}%
\providecommand \EOS [0]{\spacefactor3000\relax}%
\providecommand \BibitemShut  [1]{\csname bibitem#1\endcsname}%
\let\auto@bib@innerbib\@empty
\bibitem [{\citenamefont {Polkovnikov}\ \emph {et~al.}(2011)\citenamefont
  {Polkovnikov}, \citenamefont {Sengupta}, \citenamefont {Silva},\ and\
  \citenamefont {Vengalattore}}]{Polkovnikov}%
  \BibitemOpen
  \bibfield  {author} {\bibinfo {author} {\bibfnamefont {A.}~\bibnamefont
  {Polkovnikov}}, \bibinfo {author} {\bibfnamefont {K.}~\bibnamefont
  {Sengupta}}, \bibinfo {author} {\bibfnamefont {A.}~\bibnamefont {Silva}},\
  and\ \bibinfo {author} {\bibfnamefont {M.}~\bibnamefont {Vengalattore}},\
  }\bibfield  {title} {\bibinfo {title} {Nonequilibrium dynamics of closed
  interacting quantum systems},\ }\href
  {https://doi.org/https://doi.org/10.1103/RevModPhys.83.863} {\bibfield
  {journal} {\bibinfo  {journal} {Rev. Mod. Phys.}\ }\textbf {\bibinfo {volume}
  {83}},\ \bibinfo {pages} {863} (\bibinfo {year} {2011})}\BibitemShut
  {NoStop}%
\bibitem [{\citenamefont {Cazalilla}\ and\ \citenamefont
  {Rigol}(2010)}]{Cazalilla}%
  \BibitemOpen
  \bibfield  {author} {\bibinfo {author} {\bibfnamefont {M.~A.}\ \bibnamefont
  {Cazalilla}}\ and\ \bibinfo {author} {\bibfnamefont {M.}~\bibnamefont
  {Rigol}},\ }\bibfield  {title} {\bibinfo {title} {Focus on nonequilibrium
  dynamics of isolated one-dimensional quantum systems},\ }\href
  {https://doi.org/https://doi.org/10.1088/1367-2630/12/5/055006} {\bibfield
  {journal} {\bibinfo  {journal} {New Journal of Physics}\ }\textbf {\bibinfo
  {volume} {12}},\ \bibinfo {pages} {055006} (\bibinfo {year}
  {2010})}\BibitemShut {NoStop}%
\bibitem [{\citenamefont {Heyl}(2018)}]{Heyl2018}%
  \BibitemOpen
  \bibfield  {author} {\bibinfo {author} {\bibfnamefont {M.}~\bibnamefont
  {Heyl}},\ }\bibfield  {title} {\bibinfo {title} {Dynamical quantum phase
  transitions: A review},\ }\href
  {https://doi.org/https://doi.org/10.1088/1361-6633/aaaf9a} {\bibfield
  {journal} {\bibinfo  {journal} {Reports on Progress in Physics}\ }\textbf
  {\bibinfo {volume} {81}},\ \bibinfo {pages} {054001} (\bibinfo {year}
  {2018})}\BibitemShut {NoStop}%
\bibitem [{\citenamefont {H\"app\"ol\"a}\ \emph {et~al.}(2012)\citenamefont
  {H\"app\"ol\"a}, \citenamefont {Hal\'asz},\ and\ \citenamefont
  {Hamma}}]{Happola}%
  \BibitemOpen
  \bibfield  {author} {\bibinfo {author} {\bibfnamefont {J.}~\bibnamefont
  {H\"app\"ol\"a}}, \bibinfo {author} {\bibfnamefont {G.~B.}\ \bibnamefont
  {Hal\'asz}},\ and\ \bibinfo {author} {\bibfnamefont {A.}~\bibnamefont
  {Hamma}},\ }\bibfield  {title} {\bibinfo {title} {Universality and robustness
  of revivals in the transverse field $\mathrm{XY}$ model},\ }\href
  {https://doi.org/10.1103/PhysRevA.85.032114} {\bibfield  {journal} {\bibinfo
  {journal} {Phys. Rev. A}\ }\textbf {\bibinfo {volume} {85}},\ \bibinfo
  {pages} {032114} (\bibinfo {year} {2012})}\BibitemShut {NoStop}%
\bibitem [{\citenamefont {Huang}\ and\ \citenamefont {Kais}(2006)}]{Huang2006}%
  \BibitemOpen
  \bibfield  {author} {\bibinfo {author} {\bibfnamefont {Z.}~\bibnamefont
  {Huang}}\ and\ \bibinfo {author} {\bibfnamefont {S.}~\bibnamefont {Kais}},\
  }\bibfield  {title} {\bibinfo {title} {Entanglement evolution of
  one-dimensional spin systems in external magnetic fields},\ }\href
  {https://doi.org/10.1103/PhysRevA.73.022339} {\bibfield  {journal} {\bibinfo
  {journal} {Phys. Rev. A}\ }\textbf {\bibinfo {volume} {73}},\ \bibinfo
  {pages} {022339} (\bibinfo {year} {2006})}\BibitemShut {NoStop}%
\bibitem [{\citenamefont {Bayat}\ and\ \citenamefont
  {Bose}(2010)}]{Bayat:2010aa}%
  \BibitemOpen
  \bibfield  {author} {\bibinfo {author} {\bibfnamefont {A.}~\bibnamefont
  {Bayat}}\ and\ \bibinfo {author} {\bibfnamefont {S.}~\bibnamefont {Bose}},\
  }\bibfield  {title} {\bibinfo {title} {Information-transferring ability of
  the different phases of a finite xxz spin chain},\ }\href
  {https://doi.org/https://doi.org/10.1103/PhysRevA.81.012304} {\bibfield
  {journal} {\bibinfo  {journal} {Phys. Rev. A}\ }\textbf {\bibinfo {volume}
  {81}},\ \bibinfo {pages} {012304} (\bibinfo {year} {2010})}\BibitemShut
  {NoStop}%
\bibitem [{\citenamefont {Kitagawa}\ and\ \citenamefont
  {Ueda}(1993)}]{metrology1}%
  \BibitemOpen
  \bibfield  {author} {\bibinfo {author} {\bibfnamefont {M.}~\bibnamefont
  {Kitagawa}}\ and\ \bibinfo {author} {\bibfnamefont {M.}~\bibnamefont
  {Ueda}},\ }\bibfield  {title} {\bibinfo {title} {Squeezed spin states},\
  }\href {https://doi.org/https://doi.org/10.1103/PhysRevA.47.5138} {\bibfield
  {journal} {\bibinfo  {journal} {Phys. Rev. A}\ }\textbf {\bibinfo {volume}
  {47}},\ \bibinfo {pages} {5138} (\bibinfo {year} {1993})}\BibitemShut
  {NoStop}%
\bibitem [{\citenamefont {Bollinger}\ \emph {et~al.}(1996)\citenamefont
  {Bollinger}, \citenamefont {Itano}, \citenamefont {Wineland},\ and\
  \citenamefont {Heinzen}}]{Bollinger}%
  \BibitemOpen
  \bibfield  {author} {\bibinfo {author} {\bibfnamefont {J.~J.}\ \bibnamefont
  {Bollinger}}, \bibinfo {author} {\bibfnamefont {W.~M.}\ \bibnamefont
  {Itano}}, \bibinfo {author} {\bibfnamefont {D.~J.}\ \bibnamefont
  {Wineland}},\ and\ \bibinfo {author} {\bibfnamefont {D.~J.}\ \bibnamefont
  {Heinzen}},\ }\bibfield  {title} {\bibinfo {title} {Optimal frequency
  measurements with maximally correlated states},\ }\href
  {https://doi.org/https://doi.org/10.1103/PhysRevA.54.R4649} {\bibfield
  {journal} {\bibinfo  {journal} {Phys. Rev. A}\ }\textbf {\bibinfo {volume}
  {54}},\ \bibinfo {pages} {R4649} (\bibinfo {year} {1996})}\BibitemShut
  {NoStop}%
\bibitem [{\citenamefont {Bayat}(2014)}]{Bayat:2014aa}%
  \BibitemOpen
  \bibfield  {author} {\bibinfo {author} {\bibfnamefont {A.}~\bibnamefont
  {Bayat}},\ }\bibfield  {title} {\bibinfo {title} {Arbitrary perfect state
  transfer in d-level spin chains},\ }\href
  {https://doi.org/https://doi.org/10.1103/PhysRevA.89.062302} {\bibfield
  {journal} {\bibinfo  {journal} {Phys. Rev. A}\ }\textbf {\bibinfo {volume}
  {89}},\ \bibinfo {pages} {062302} (\bibinfo {year} {2014})}\BibitemShut
  {NoStop}%
\bibitem [{\citenamefont {Raussendorf}\ and\ \citenamefont
  {Briegel}(2001)}]{Raussendorf}%
  \BibitemOpen
  \bibfield  {author} {\bibinfo {author} {\bibfnamefont {R.}~\bibnamefont
  {Raussendorf}}\ and\ \bibinfo {author} {\bibfnamefont {H.~J.}\ \bibnamefont
  {Briegel}},\ }\bibfield  {title} {\bibinfo {title} {A one-way quantum
  computer},\ }\href
  {https://doi.org/https://doi.org/10.1103/PhysRevLett.86.5188} {\bibfield
  {journal} {\bibinfo  {journal} {Phys. Rev. Lett.}\ }\textbf {\bibinfo
  {volume} {86}},\ \bibinfo {pages} {5188} (\bibinfo {year}
  {2001})}\BibitemShut {NoStop}%
\bibitem [{\citenamefont {Bayat}\ and\ \citenamefont
  {Karimipour}(2007)}]{Bayat:2007aa}%
  \BibitemOpen
  \bibfield  {author} {\bibinfo {author} {\bibfnamefont {A.}~\bibnamefont
  {Bayat}}\ and\ \bibinfo {author} {\bibfnamefont {V.}~\bibnamefont
  {Karimipour}},\ }\bibfield  {title} {\bibinfo {title} {Transfer of d -level
  quantum states through spin chains by random swapping},\ }\href
  {https://doi.org/https://doi.org/10.1103/PhysRevA.75.022321} {\bibfield
  {journal} {\bibinfo  {journal} {Phys. Rev. A}\ }\textbf {\bibinfo {volume}
  {75}},\ \bibinfo {pages} {022321} (\bibinfo {year} {2007})}\BibitemShut
  {NoStop}%
\bibitem [{\citenamefont {Lamacraft}\ and\ \citenamefont
  {Moore}(2012)}]{Lamacraft}%
  \BibitemOpen
  \bibfield  {author} {\bibinfo {author} {\bibfnamefont {A.}~\bibnamefont
  {Lamacraft}}\ and\ \bibinfo {author} {\bibfnamefont {J.}~\bibnamefont
  {Moore}},\ }in\ \href
  {https://doi.org/https://doi.org/10.1016/B978-0-444-53857-4.00007-6} {\emph
  {\bibinfo {booktitle} {Ultracold Bosonic and Fermionic Gases}}},\ \bibinfo
  {series} {Contemporary Concepts of Condensed Matter Science}, Vol.~\bibinfo
  {volume} {5},\ \bibinfo {editor} {edited by\ \bibinfo {editor} {\bibfnamefont
  {K.}~\bibnamefont {Levin}}, \bibinfo {editor} {\bibfnamefont {A.~L.}\
  \bibnamefont {Fetter}},\ and\ \bibinfo {editor} {\bibfnamefont {D.~M.}\
  \bibnamefont {Stamper-Kurn}}}\ (\bibinfo  {publisher} {Elsevier},\ \bibinfo
  {year} {2012})\ pp.\ \bibinfo {pages} {177 -- 202}\BibitemShut {NoStop}%
\bibitem [{\citenamefont {Gedik}\ \emph {et~al.}(2007)\citenamefont {Gedik},
  \citenamefont {Yang}, \citenamefont {Logvenov}, \citenamefont {Bozovic},\
  and\ \citenamefont {Zewail}}]{Gedik}%
  \BibitemOpen
  \bibfield  {author} {\bibinfo {author} {\bibfnamefont {N.}~\bibnamefont
  {Gedik}}, \bibinfo {author} {\bibfnamefont {D.-S.}\ \bibnamefont {Yang}},
  \bibinfo {author} {\bibfnamefont {G.}~\bibnamefont {Logvenov}}, \bibinfo
  {author} {\bibfnamefont {I.}~\bibnamefont {Bozovic}},\ and\ \bibinfo {author}
  {\bibfnamefont {A.~H.}\ \bibnamefont {Zewail}},\ }\bibfield  {title}
  {\bibinfo {title} {Nonequilibrium phase transitions in cuprates observed by
  ultrafast electron crystallography},\ }\href
  {https://doi.org/https://doi.org/10.1126/science.1138834} {\bibfield
  {journal} {\bibinfo  {journal} {Science}\ }\textbf {\bibinfo {volume}
  {316}},\ \bibinfo {pages} {425} (\bibinfo {year} {2007})}\BibitemShut
  {NoStop}%
\bibitem [{\citenamefont {Mandel}\ \emph {et~al.}(2003)\citenamefont {Mandel},
  \citenamefont {Greiner}, \citenamefont {Widera}, \citenamefont {Rom},
  \citenamefont {H\"ansch},\ and\ \citenamefont {Bloch}}]{Mandel}%
  \BibitemOpen
  \bibfield  {author} {\bibinfo {author} {\bibfnamefont {O.}~\bibnamefont
  {Mandel}}, \bibinfo {author} {\bibfnamefont {M.}~\bibnamefont {Greiner}},
  \bibinfo {author} {\bibfnamefont {A.}~\bibnamefont {Widera}}, \bibinfo
  {author} {\bibfnamefont {T.}~\bibnamefont {Rom}}, \bibinfo {author}
  {\bibfnamefont {T.~W.}\ \bibnamefont {H\"ansch}},\ and\ \bibinfo {author}
  {\bibfnamefont {I.}~\bibnamefont {Bloch}},\ }\bibfield  {title} {\bibinfo
  {title} {Controlled collisions for multi-particle entanglement of optically
  trapped atoms},\ }\href {https://doi.org/https://doi.org/10.1038/nature02008}
  {\bibfield  {journal} {\bibinfo  {journal} {Nature}\ }\textbf {\bibinfo
  {volume} {425}},\ \bibinfo {pages} {937} (\bibinfo {year}
  {2003})}\BibitemShut {NoStop}%
\bibitem [{\citenamefont {Bloch}(2005)}]{Bloch:2005aa}%
  \BibitemOpen
  \bibfield  {author} {\bibinfo {author} {\bibfnamefont {I.}~\bibnamefont
  {Bloch}},\ }\bibfield  {title} {\bibinfo {title} {Exploring quantum matter
  with ultracold atoms in optical lattices},\ }\href
  {https://doi.org/10.1088/0953-4075/38/9/013} {\bibfield  {journal} {\bibinfo
  {journal} {Journal of Physics B: Atomic, Molecular and Optical Physics}\
  }\textbf {\bibinfo {volume} {38}},\ \bibinfo {pages} {S629} (\bibinfo {year}
  {2005})}\BibitemShut {NoStop}%
\bibitem [{\citenamefont {Treutlein}\ \emph {et~al.}(2006)\citenamefont
  {Treutlein}, \citenamefont {Steinmetz}, \citenamefont {Colombe},
  \citenamefont {Lev}, \citenamefont {Hommelhoff}, \citenamefont {Reichel},
  \citenamefont {Greiner}, \citenamefont {Mandel}, \citenamefont {Widera},
  \citenamefont {Rom}, \citenamefont {Bloch},\ and\ \citenamefont
  {H{\"a}nsch}}]{Treutlein}%
  \BibitemOpen
  \bibfield  {author} {\bibinfo {author} {\bibfnamefont {P.}~\bibnamefont
  {Treutlein}}, \bibinfo {author} {\bibfnamefont {T.}~\bibnamefont
  {Steinmetz}}, \bibinfo {author} {\bibfnamefont {Y.}~\bibnamefont {Colombe}},
  \bibinfo {author} {\bibfnamefont {B.}~\bibnamefont {Lev}}, \bibinfo {author}
  {\bibfnamefont {P.}~\bibnamefont {Hommelhoff}}, \bibinfo {author}
  {\bibfnamefont {J.}~\bibnamefont {Reichel}}, \bibinfo {author} {\bibfnamefont
  {M.}~\bibnamefont {Greiner}}, \bibinfo {author} {\bibfnamefont
  {O.}~\bibnamefont {Mandel}}, \bibinfo {author} {\bibfnamefont
  {A.}~\bibnamefont {Widera}}, \bibinfo {author} {\bibfnamefont
  {T.}~\bibnamefont {Rom}}, \bibinfo {author} {\bibfnamefont {I.}~\bibnamefont
  {Bloch}},\ and\ \bibinfo {author} {\bibfnamefont {T.}~\bibnamefont
  {H{\"a}nsch}},\ }\bibfield  {title} {\bibinfo {title} {Quantum information
  processing in optical lattices and magnetic microtraps},\ }\href
  {https://doi.org/https://doi.org/10.1002/prop.200610325} {\bibfield
  {journal} {\bibinfo  {journal} {Fortschritte der Physik}\ }\textbf {\bibinfo
  {volume} {54}},\ \bibinfo {pages} {702} (\bibinfo {year} {2006})}\BibitemShut
  {NoStop}%
\bibitem [{\citenamefont {Cramer}\ \emph {et~al.}(2013)\citenamefont {Cramer},
  \citenamefont {Bernard}, \citenamefont {Fabbri}, \citenamefont {Fallani},
  \citenamefont {Fort}, \citenamefont {Rosi}, \citenamefont {Caruso},
  \citenamefont {Inguscio},\ and\ \citenamefont {Plenio}}]{Cramer:2013aa}%
  \BibitemOpen
  \bibfield  {author} {\bibinfo {author} {\bibfnamefont {M.}~\bibnamefont
  {Cramer}}, \bibinfo {author} {\bibfnamefont {A.}~\bibnamefont {Bernard}},
  \bibinfo {author} {\bibfnamefont {N.}~\bibnamefont {Fabbri}}, \bibinfo
  {author} {\bibfnamefont {L.}~\bibnamefont {Fallani}}, \bibinfo {author}
  {\bibfnamefont {C.}~\bibnamefont {Fort}}, \bibinfo {author} {\bibfnamefont
  {S.}~\bibnamefont {Rosi}}, \bibinfo {author} {\bibfnamefont {F.}~\bibnamefont
  {Caruso}}, \bibinfo {author} {\bibfnamefont {M.}~\bibnamefont {Inguscio}},\
  and\ \bibinfo {author} {\bibfnamefont {M.~B.}\ \bibnamefont {Plenio}},\
  }\bibfield  {title} {\bibinfo {title} {Spatial entanglement of bosons in
  optical lattices},\ }\href
  {https://doi.org/https://doi.org/10.1038/ncomms3161} {\bibfield  {journal}
  {\bibinfo  {journal} {Nature Communications}\ }\textbf {\bibinfo {volume}
  {4}},\ \bibinfo {pages} {2161} (\bibinfo {year} {2013})}\BibitemShut
  {NoStop}%
\bibitem [{\citenamefont {Leibfried}\ \emph {et~al.}(2003)\citenamefont
  {Leibfried}, \citenamefont {Blatt}, \citenamefont {Monroe},\ and\
  \citenamefont {Wineland}}]{Leibfried}%
  \BibitemOpen
  \bibfield  {author} {\bibinfo {author} {\bibfnamefont {D.}~\bibnamefont
  {Leibfried}}, \bibinfo {author} {\bibfnamefont {R.}~\bibnamefont {Blatt}},
  \bibinfo {author} {\bibfnamefont {C.}~\bibnamefont {Monroe}},\ and\ \bibinfo
  {author} {\bibfnamefont {D.}~\bibnamefont {Wineland}},\ }\bibfield  {title}
  {\bibinfo {title} {Quantum dynamics of single trapped ions},\ }\href
  {https://doi.org/https://doi.org/10.1103/RevModPhys.75.281} {\bibfield
  {journal} {\bibinfo  {journal} {Rev. Mod. Phys.}\ }\textbf {\bibinfo {volume}
  {75}},\ \bibinfo {pages} {281} (\bibinfo {year} {2003})}\BibitemShut
  {NoStop}%
\bibitem [{\citenamefont {Jafari}\ and\ \citenamefont
  {Akbari}(2015)}]{Jafari2015}%
  \BibitemOpen
  \bibfield  {author} {\bibinfo {author} {\bibfnamefont {R.}~\bibnamefont
  {Jafari}}\ and\ \bibinfo {author} {\bibfnamefont {A.}~\bibnamefont
  {Akbari}},\ }\bibfield  {title} {\bibinfo {title} {Gapped quantum criticality
  gains long-time quantum correlations},\ }\href
  {https://doi.org/10.1209/0295-5075/111/10007} {\bibfield  {journal} {\bibinfo
   {journal} {Europhysics Letters}\ }\textbf {\bibinfo {volume} {111}},\
  \bibinfo {pages} {10007} (\bibinfo {year} {2015})}\BibitemShut {NoStop}%
\bibitem [{\citenamefont {Jafari}(2016)}]{Jafari2016}%
  \BibitemOpen
  \bibfield  {author} {\bibinfo {author} {\bibfnamefont {R.}~\bibnamefont
  {Jafari}},\ }\bibfield  {title} {\bibinfo {title} {Quench dynamics and ground
  state fidelity of the one-dimensional extended quantum compass model in a
  transverse field},\ }\href {https://doi.org/10.1088/1751-8113/49/18/185004}
  {\bibfield  {journal} {\bibinfo  {journal} {Journal of Physics A:
  Mathematical and Theoretical}\ }\textbf {\bibinfo {volume} {49}},\ \bibinfo
  {pages} {185004} (\bibinfo {year} {2016})}\BibitemShut {NoStop}%
\bibitem [{\citenamefont {Sharma}\ \emph {et~al.}(2015)\citenamefont {Sharma},
  \citenamefont {Suzuki},\ and\ \citenamefont {Dutta}}]{Sharma2015}%
  \BibitemOpen
  \bibfield  {author} {\bibinfo {author} {\bibfnamefont {S.}~\bibnamefont
  {Sharma}}, \bibinfo {author} {\bibfnamefont {S.}~\bibnamefont {Suzuki}},\
  and\ \bibinfo {author} {\bibfnamefont {A.}~\bibnamefont {Dutta}},\ }\bibfield
   {title} {\bibinfo {title} {Quenches and dynamical phase transitions in a
  nonintegrable quantum ising model},\ }\href
  {https://doi.org/10.1103/PhysRevB.92.104306} {\bibfield  {journal} {\bibinfo
  {journal} {Phys. Rev. B}\ }\textbf {\bibinfo {volume} {92}},\ \bibinfo
  {pages} {104306} (\bibinfo {year} {2015})}\BibitemShut {NoStop}%
\bibitem [{\citenamefont {Montes}\ and\ \citenamefont
  {Hamma}(2012)}]{Montes2012}%
  \BibitemOpen
  \bibfield  {author} {\bibinfo {author} {\bibfnamefont {S.}~\bibnamefont
  {Montes}}\ and\ \bibinfo {author} {\bibfnamefont {A.}~\bibnamefont {Hamma}},\
  }\bibfield  {title} {\bibinfo {title} {Phase diagram and quench dynamics of
  the cluster- $\mathrm{XY}$ spin chain},\ }\href
  {https://doi.org/https://doi.org/10.1103/PhysRevE.86.021101} {\bibfield
  {journal} {\bibinfo  {journal} {Phys. Rev. E}\ }\textbf {\bibinfo {volume}
  {86}},\ \bibinfo {pages} {021101} (\bibinfo {year} {2012})}\BibitemShut
  {NoStop}%
\bibitem [{\citenamefont {Sacramento}(2014)}]{Sacramento2014}%
  \BibitemOpen
  \bibfield  {author} {\bibinfo {author} {\bibfnamefont {P.~D.}\ \bibnamefont
  {Sacramento}},\ }\bibfield  {title} {\bibinfo {title} {Fate of majorana
  fermions and chern numbers after a quantum quench},\ }\href
  {https://doi.org/10.1103/PhysRevE.90.032138} {\bibfield  {journal} {\bibinfo
  {journal} {Phys. Rev. E}\ }\textbf {\bibinfo {volume} {90}},\ \bibinfo
  {pages} {032138} (\bibinfo {year} {2014})}\BibitemShut {NoStop}%
\bibitem [{\citenamefont {Jafari}\ and\ \citenamefont
  {Johannesson}(2017)}]{Jafari2017}%
  \BibitemOpen
  \bibfield  {author} {\bibinfo {author} {\bibfnamefont {R.}~\bibnamefont
  {Jafari}}\ and\ \bibinfo {author} {\bibfnamefont {H.}~\bibnamefont
  {Johannesson}},\ }\bibfield  {title} {\bibinfo {title} {Loschmidt echo
  revivals: Critical and noncritical},\ }\href
  {https://doi.org/10.1103/PhysRevLett.118.015701} {\bibfield  {journal}
  {\bibinfo  {journal} {Phys. Rev. Lett.}\ }\textbf {\bibinfo {volume} {118}},\
  \bibinfo {pages} {015701} (\bibinfo {year} {2017})}\BibitemShut {NoStop}%
\bibitem [{\citenamefont {Greiner}\ \emph {et~al.}(2002)\citenamefont
  {Greiner}, \citenamefont {Mandel}, \citenamefont {H\"ansch},\ and\
  \citenamefont {Bloch}}]{Greiner2002}%
  \BibitemOpen
  \bibfield  {author} {\bibinfo {author} {\bibfnamefont {M.}~\bibnamefont
  {Greiner}}, \bibinfo {author} {\bibfnamefont {O.}~\bibnamefont {Mandel}},
  \bibinfo {author} {\bibfnamefont {T.~W.}\ \bibnamefont {H\"ansch}},\ and\
  \bibinfo {author} {\bibfnamefont {I.}~\bibnamefont {Bloch}},\ }\bibfield
  {title} {\bibinfo {title} {Collapse and revival of the matter wave field of a
  bose-einstein condensate},\ }\href
  {https://doi.org/https://doi.org/10.1038/nature00968} {\bibfield  {journal}
  {\bibinfo  {journal} {Nature}\ }\textbf {\bibinfo {volume} {419}},\ \bibinfo
  {pages} {51} (\bibinfo {year} {2002})}\BibitemShut {NoStop}%
\bibitem [{\citenamefont {Calabrese}\ and\ \citenamefont
  {Cardy}(2006)}]{Calabrese2006}%
  \BibitemOpen
  \bibfield  {author} {\bibinfo {author} {\bibfnamefont {P.}~\bibnamefont
  {Calabrese}}\ and\ \bibinfo {author} {\bibfnamefont {J.}~\bibnamefont
  {Cardy}},\ }\bibfield  {title} {\bibinfo {title} {Time dependence of
  correlation functions following a quantum quench},\ }\href
  {https://doi.org/https://doi.org/10.1103/PhysRevLett.96.136801} {\bibfield
  {journal} {\bibinfo  {journal} {Phys. Rev. Lett.}\ }\textbf {\bibinfo
  {volume} {96}},\ \bibinfo {pages} {136801} (\bibinfo {year}
  {2006})}\BibitemShut {NoStop}%
\bibitem [{\citenamefont {Mitra}(2018)}]{Mitra:2018aa}%
  \BibitemOpen
  \bibfield  {author} {\bibinfo {author} {\bibfnamefont {A.}~\bibnamefont
  {Mitra}},\ }\bibfield  {title} {\bibinfo {title} {Quantum quench dynamics},\
  }\href {https://doi.org/10.1146/annurev-conmatphys-031016-025451} {\bibfield
  {journal} {\bibinfo  {journal} {Annual Review of Condensed Matter Physics}\
  }\textbf {\bibinfo {volume} {9}},\ \bibinfo {pages} {245} (\bibinfo {year}
  {2018})}\BibitemShut {NoStop}%
\bibitem [{\citenamefont {Nag}\ \emph {et~al.}(2013)\citenamefont {Nag},
  \citenamefont {Dutta},\ and\ \citenamefont {Patra}}]{NAG:2013aa}%
  \BibitemOpen
  \bibfield  {author} {\bibinfo {author} {\bibfnamefont {T.}~\bibnamefont
  {Nag}}, \bibinfo {author} {\bibfnamefont {A.}~\bibnamefont {Dutta}},\ and\
  \bibinfo {author} {\bibfnamefont {A.}~\bibnamefont {Patra}},\ }\bibfield
  {title} {\bibinfo {title} {Quenching dynamics and quantum information},\
  }\href {https://doi.org/10.1142/S0217979213450367} {\bibfield  {journal}
  {\bibinfo  {journal} {International Journal of Modern Physics B}\ }\textbf
  {\bibinfo {volume} {27}},\ \bibinfo {pages} {1345036} (\bibinfo {year}
  {2013})}\BibitemShut {NoStop}%
\bibitem [{\citenamefont {Gorin}\ \emph {et~al.}(2006)\citenamefont {Gorin},
  \citenamefont {Prosen}, \citenamefont {Seligman},\ and\ \citenamefont {{\v
  Z}nidari{\v c}}}]{Gorin}%
  \BibitemOpen
  \bibfield  {author} {\bibinfo {author} {\bibfnamefont {T.}~\bibnamefont
  {Gorin}}, \bibinfo {author} {\bibfnamefont {T.}~\bibnamefont {Prosen}},
  \bibinfo {author} {\bibfnamefont {T.~H.}\ \bibnamefont {Seligman}},\ and\
  \bibinfo {author} {\bibfnamefont {M.}~\bibnamefont {{\v Z}nidari{\v c}}},\
  }\bibfield  {title} {\bibinfo {title} {Dynamics of loschmidt echoes and
  fidelity decay},\ }\href
  {https://doi.org/https://doi.org/10.1016/j.physrep.2006.09.003} {\bibfield
  {journal} {\bibinfo  {journal} {Physics Reports}\ }\textbf {\bibinfo {volume}
  {435}},\ \bibinfo {pages} {33} (\bibinfo {year} {2006})}\BibitemShut
  {NoStop}%
\bibitem [{\citenamefont {Jacquod}\ and\ \citenamefont
  {Petitjean}(2009)}]{Jacquod}%
  \BibitemOpen
  \bibfield  {author} {\bibinfo {author} {\bibfnamefont {P.}~\bibnamefont
  {Jacquod}}\ and\ \bibinfo {author} {\bibfnamefont {C.}~\bibnamefont
  {Petitjean}},\ }\bibfield  {title} {\bibinfo {title} {Decoherence,
  entanglement and irreversibility in quantum dynamical systems with few
  degrees of freedom},\ }\href {https://doi.org/10.1080/00018730902831009}
  {\bibfield  {journal} {\bibinfo  {journal} {Advances in Physics}\ }\textbf
  {\bibinfo {volume} {58}},\ \bibinfo {pages} {67} (\bibinfo {year}
  {2009})}\BibitemShut {NoStop}%
\bibitem [{\citenamefont {Kolodrubetz}\ \emph {et~al.}(2012)\citenamefont
  {Kolodrubetz}, \citenamefont {Clark},\ and\ \citenamefont {Huse}}]{Clark}%
  \BibitemOpen
  \bibfield  {author} {\bibinfo {author} {\bibfnamefont {M.}~\bibnamefont
  {Kolodrubetz}}, \bibinfo {author} {\bibfnamefont {B.~K.}\ \bibnamefont
  {Clark}},\ and\ \bibinfo {author} {\bibfnamefont {D.~A.}\ \bibnamefont
  {Huse}},\ }\bibfield  {title} {\bibinfo {title} {Nonequilibrium dynamic
  critical scaling of the quantum ising chain},\ }\href
  {https://doi.org/10.1103/PhysRevLett.109.015701} {\bibfield  {journal}
  {\bibinfo  {journal} {Phys. Rev. Lett.}\ }\textbf {\bibinfo {volume} {109}},\
  \bibinfo {pages} {015701} (\bibinfo {year} {2012})}\BibitemShut {NoStop}%
\bibitem [{\citenamefont {Zurek}\ \emph {et~al.}(2005)\citenamefont {Zurek},
  \citenamefont {Dorner},\ and\ \citenamefont {Zoller}}]{Zurek}%
  \BibitemOpen
  \bibfield  {author} {\bibinfo {author} {\bibfnamefont {W.~H.}\ \bibnamefont
  {Zurek}}, \bibinfo {author} {\bibfnamefont {U.}~\bibnamefont {Dorner}},\ and\
  \bibinfo {author} {\bibfnamefont {P.}~\bibnamefont {Zoller}},\ }\bibfield
  {title} {\bibinfo {title} {Dynamics of a quantum phase transition},\ }\href
  {https://doi.org/https://doi.org/10.1103/PhysRevLett.95.105701} {\bibfield
  {journal} {\bibinfo  {journal} {Phys. Rev. Lett.}\ }\textbf {\bibinfo
  {volume} {95}},\ \bibinfo {pages} {105701} (\bibinfo {year}
  {2005})}\BibitemShut {NoStop}%
\bibitem [{\citenamefont {Heyl}\ \emph {et~al.}(2013)\citenamefont {Heyl},
  \citenamefont {Polkovnikov},\ and\ \citenamefont {Kehrein}}]{Heyl2013}%
  \BibitemOpen
  \bibfield  {author} {\bibinfo {author} {\bibfnamefont {M.}~\bibnamefont
  {Heyl}}, \bibinfo {author} {\bibfnamefont {A.}~\bibnamefont {Polkovnikov}},\
  and\ \bibinfo {author} {\bibfnamefont {S.}~\bibnamefont {Kehrein}},\
  }\bibfield  {title} {\bibinfo {title} {Dynamical quantum phase transitions in
  the transverse-field ising model},\ }\href
  {https://doi.org/https://doi.org/10.1103/PhysRevLett.110.135704} {\bibfield
  {journal} {\bibinfo  {journal} {Phys. Rev. Lett.}\ }\textbf {\bibinfo
  {volume} {110}},\ \bibinfo {pages} {135704} (\bibinfo {year}
  {2013})}\BibitemShut {NoStop}%
\bibitem [{\citenamefont {Amico}\ \emph {et~al.}(2008)\citenamefont {Amico},
  \citenamefont {Fazio}, \citenamefont {Osterloh},\ and\ \citenamefont
  {Vedral}}]{Fazio2008}%
  \BibitemOpen
  \bibfield  {author} {\bibinfo {author} {\bibfnamefont {L.}~\bibnamefont
  {Amico}}, \bibinfo {author} {\bibfnamefont {R.}~\bibnamefont {Fazio}},
  \bibinfo {author} {\bibfnamefont {A.}~\bibnamefont {Osterloh}},\ and\
  \bibinfo {author} {\bibfnamefont {V.}~\bibnamefont {Vedral}},\ }\bibfield
  {title} {\bibinfo {title} {Entanglement in many-body systems},\ }\href
  {https://doi.org/https://doi.org/10.1103/RevModPhys.80.517} {\bibfield
  {journal} {\bibinfo  {journal} {Rev. Mod. Phys.}\ }\textbf {\bibinfo {volume}
  {80}},\ \bibinfo {pages} {517} (\bibinfo {year} {2008})}\BibitemShut
  {NoStop}%
\bibitem [{\citenamefont {Eisert}\ \emph {et~al.}(2010)\citenamefont {Eisert},
  \citenamefont {Cramer},\ and\ \citenamefont {Plenio}}]{Eisert:2010aa}%
  \BibitemOpen
  \bibfield  {author} {\bibinfo {author} {\bibfnamefont {J.}~\bibnamefont
  {Eisert}}, \bibinfo {author} {\bibfnamefont {M.}~\bibnamefont {Cramer}},\
  and\ \bibinfo {author} {\bibfnamefont {M.~B.}\ \bibnamefont {Plenio}},\
  }\bibfield  {title} {\bibinfo {title} {Colloquium: Area laws for the
  entanglement entropy},\ }\href {https://doi.org/10.1103/RevModPhys.82.277}
  {\bibfield  {journal} {\bibinfo  {journal} {Rev. Mod. Phys.}\ }\textbf
  {\bibinfo {volume} {82}},\ \bibinfo {pages} {277} (\bibinfo {year}
  {2010})}\BibitemShut {NoStop}%
\bibitem [{\citenamefont {Plenio}\ and\ \citenamefont {Huelga}(2008)}]{Plenio}%
  \BibitemOpen
  \bibfield  {author} {\bibinfo {author} {\bibfnamefont {M.~B.}\ \bibnamefont
  {Plenio}}\ and\ \bibinfo {author} {\bibfnamefont {S.~F.}\ \bibnamefont
  {Huelga}},\ }\bibfield  {title} {\bibinfo {title} {Dephasing-assisted
  transport: Quantum networks and biomolecules},\ }\href
  {https://doi.org/10.1088/1367-2630/10/11/113019} {\bibfield  {journal}
  {\bibinfo  {journal} {New J. Phys.}\ }\textbf {\bibinfo {volume} {10}},\
  \bibinfo {pages} {113019} (\bibinfo {year} {2008})}\BibitemShut {NoStop}%
\bibitem [{\citenamefont {Rebentrost}\ \emph {et~al.}(2009)\citenamefont
  {Rebentrost}, \citenamefont {Mohseni},\ and\ \citenamefont
  {Aspuru-Guzik}}]{Rebentrost}%
  \BibitemOpen
  \bibfield  {author} {\bibinfo {author} {\bibfnamefont {P.}~\bibnamefont
  {Rebentrost}}, \bibinfo {author} {\bibfnamefont {M.}~\bibnamefont
  {Mohseni}},\ and\ \bibinfo {author} {\bibfnamefont {A.}~\bibnamefont
  {Aspuru-Guzik}},\ }\bibfield  {title} {\bibinfo {title} {Role of quantum
  coherence and environmental fluctuations in chromophoric energy transport},\
  }\href {https://doi.org/10.1021/jp901724d} {\bibfield  {journal} {\bibinfo
  {journal} {The Journal of Physical Chemistry B}\ }\textbf {\bibinfo {volume}
  {113}},\ \bibinfo {pages} {9942} (\bibinfo {year} {2009})}\BibitemShut
  {NoStop}%
\bibitem [{\citenamefont {Lloyd}(2011)}]{Lloyd}%
  \BibitemOpen
  \bibfield  {author} {\bibinfo {author} {\bibfnamefont {S.}~\bibnamefont
  {Lloyd}},\ }\bibfield  {title} {\bibinfo {title} {Quantum coherence in
  biological systems},\ }\href {https://doi.org/10.1088/1742-6596/302/1/012037}
  {\bibfield  {journal} {\bibinfo  {journal} {Journal of Physics: Conference
  Series}\ }\textbf {\bibinfo {volume} {302}},\ \bibinfo {pages} {012037}
  (\bibinfo {year} {2011})}\BibitemShut {NoStop}%
\bibitem [{\citenamefont {Huelga}\ and\ \citenamefont {Plenio}(2013)}]{Huelga}%
  \BibitemOpen
  \bibfield  {author} {\bibinfo {author} {\bibfnamefont {S.}~\bibnamefont
  {Huelga}}\ and\ \bibinfo {author} {\bibfnamefont {M.}~\bibnamefont
  {Plenio}},\ }\bibfield  {title} {\bibinfo {title} {Vibrations, quanta and
  biology},\ }\href {https://doi.org/10.1080/00405000.2013.829687} {\bibfield
  {journal} {\bibinfo  {journal} {Contemporary Physics}\ }\textbf {\bibinfo
  {volume} {54}},\ \bibinfo {pages} {181} (\bibinfo {year} {2013})}\BibitemShut
  {NoStop}%
\bibitem [{\citenamefont {Li}\ \emph {et~al.}(2012)\citenamefont {Li},
  \citenamefont {Lambert}, \citenamefont {Chen}, \citenamefont {Chen},\ and\
  \citenamefont {Nori}}]{Li2012}%
  \BibitemOpen
  \bibfield  {author} {\bibinfo {author} {\bibfnamefont {C.-M.}\ \bibnamefont
  {Li}}, \bibinfo {author} {\bibfnamefont {N.}~\bibnamefont {Lambert}},
  \bibinfo {author} {\bibfnamefont {Y.-N.}\ \bibnamefont {Chen}}, \bibinfo
  {author} {\bibfnamefont {G.-Y.}\ \bibnamefont {Chen}},\ and\ \bibinfo
  {author} {\bibfnamefont {F.}~\bibnamefont {Nori}},\ }\bibfield  {title}
  {\bibinfo {title} {Witnessing quantum coherence: From solid-state to
  biological systems},\ }\href {https://doi.org/10.1038/srep00885} {\bibfield
  {journal} {\bibinfo  {journal} {Sci. Rep.}\ }\textbf {\bibinfo {volume}
  {2}},\ \bibinfo {pages} {885} (\bibinfo {year} {2012})}\BibitemShut {NoStop}%
\bibitem [{\citenamefont {Lambert}\ \emph {et~al.}(2013)\citenamefont
  {Lambert}, \citenamefont {Chen}, \citenamefont {Cheng}, \citenamefont {Li},
  \citenamefont {Chen},\ and\ \citenamefont {Nori}}]{Lambert2013}%
  \BibitemOpen
  \bibfield  {author} {\bibinfo {author} {\bibfnamefont {N.}~\bibnamefont
  {Lambert}}, \bibinfo {author} {\bibfnamefont {Y.-N.}\ \bibnamefont {Chen}},
  \bibinfo {author} {\bibfnamefont {Y.-C.}\ \bibnamefont {Cheng}}, \bibinfo
  {author} {\bibfnamefont {C.-M.}\ \bibnamefont {Li}}, \bibinfo {author}
  {\bibfnamefont {G.-Y.}\ \bibnamefont {Chen}},\ and\ \bibinfo {author}
  {\bibfnamefont {F.}~\bibnamefont {Nori}},\ }\bibfield  {title} {\bibinfo
  {title} {Quantum biology},\ }\href
  {https://doi.org/https://doi.org/10.1038/nphys2474} {\bibfield  {journal}
  {\bibinfo  {journal} {Nat. Phys.}\ }\textbf {\bibinfo {volume} {9}},\
  \bibinfo {pages} {10} (\bibinfo {year} {2013})}\BibitemShut {NoStop}%
\bibitem [{\citenamefont {Narasimhachar}\ and\ \citenamefont
  {Gour}(2015)}]{Narasimhachar}%
  \BibitemOpen
  \bibfield  {author} {\bibinfo {author} {\bibfnamefont {V.}~\bibnamefont
  {Narasimhachar}}\ and\ \bibinfo {author} {\bibfnamefont {G.}~\bibnamefont
  {Gour}},\ }\bibfield  {title} {\bibinfo {title} {Low-temperature
  thermodynamics with quantum coherence},\ }\href
  {https://doi.org/10.1038/ncomms8689} {\bibfield  {journal} {\bibinfo
  {journal} {Nature Communications}\ }\textbf {\bibinfo {volume} {6}},\
  \bibinfo {pages} {7689} (\bibinfo {year} {2015})}\BibitemShut {NoStop}%
\bibitem [{\citenamefont {Lostaglio}\ \emph
  {et~al.}(2015{\natexlab{a}})\citenamefont {Lostaglio}, \citenamefont
  {Jennings},\ and\ \citenamefont {Rudolph}}]{Lostaglio1}%
  \BibitemOpen
  \bibfield  {author} {\bibinfo {author} {\bibfnamefont {M.}~\bibnamefont
  {Lostaglio}}, \bibinfo {author} {\bibfnamefont {D.}~\bibnamefont
  {Jennings}},\ and\ \bibinfo {author} {\bibfnamefont {T.}~\bibnamefont
  {Rudolph}},\ }\bibfield  {title} {\bibinfo {title} {Description of quantum
  coherence in thermodynamic processes requires constraints beyond free
  energy},\ }\href {https://doi.org/https://doi.org/10.1038/ncomms7383}
  {\bibfield  {journal} {\bibinfo  {journal} {Nat. Commun.}\ }\textbf {\bibinfo
  {volume} {6}},\ \bibinfo {pages} {6383} (\bibinfo {year}
  {2015}{\natexlab{a}})}\BibitemShut {NoStop}%
\bibitem [{\citenamefont {Lostaglio}\ \emph
  {et~al.}(2015{\natexlab{b}})\citenamefont {Lostaglio}, \citenamefont
  {Korzekwa}, \citenamefont {Jennings},\ and\ \citenamefont
  {Rudolph}}]{Lostaglio2}%
  \BibitemOpen
  \bibfield  {author} {\bibinfo {author} {\bibfnamefont {M.}~\bibnamefont
  {Lostaglio}}, \bibinfo {author} {\bibfnamefont {K.}~\bibnamefont {Korzekwa}},
  \bibinfo {author} {\bibfnamefont {D.}~\bibnamefont {Jennings}},\ and\
  \bibinfo {author} {\bibfnamefont {T.}~\bibnamefont {Rudolph}},\ }\bibfield
  {title} {\bibinfo {title} {Quantum coherence, time-translation symmetry, and
  thermodynamics},\ }\href
  {https://doi.org/https://doi.org/10.1103/PhysRevX.5.021001} {\bibfield
  {journal} {\bibinfo  {journal} {Phys. Rev. X}\ }\textbf {\bibinfo {volume}
  {5}},\ \bibinfo {pages} {021001} (\bibinfo {year}
  {2015}{\natexlab{b}})}\BibitemShut {NoStop}%
\bibitem [{\citenamefont {Gour}\ \emph {et~al.}(2015)\citenamefont {Gour},
  \citenamefont {M{\"u}ller}, \citenamefont {Narasimhachar}, \citenamefont
  {Spekkens},\ and\ \citenamefont {{Yunger Halpern}}}]{Gour}%
  \BibitemOpen
  \bibfield  {author} {\bibinfo {author} {\bibfnamefont {G.}~\bibnamefont
  {Gour}}, \bibinfo {author} {\bibfnamefont {M.~P.}\ \bibnamefont
  {M{\"u}ller}}, \bibinfo {author} {\bibfnamefont {V.}~\bibnamefont
  {Narasimhachar}}, \bibinfo {author} {\bibfnamefont {R.~W.}\ \bibnamefont
  {Spekkens}},\ and\ \bibinfo {author} {\bibfnamefont {N.}~\bibnamefont
  {{Yunger Halpern}}},\ }\bibfield  {title} {\bibinfo {title} {The resource
  theory of informational nonequilibrium in thermodynamics},\ }\href
  {https://doi.org/https://doi.org/10.1016/j.physrep.2015.04.003} {\bibfield
  {journal} {\bibinfo  {journal} {Physics Reports}\ }\textbf {\bibinfo {volume}
  {583}},\ \bibinfo {pages} {1} (\bibinfo {year} {2015})}\BibitemShut {NoStop}%
\bibitem [{\citenamefont {{\AA}berg}(2014)}]{Aberg}%
  \BibitemOpen
  \bibfield  {author} {\bibinfo {author} {\bibfnamefont {J.}~\bibnamefont
  {{\AA}berg}},\ }\bibfield  {title} {\bibinfo {title} {Catalytic coherence},\
  }\href {https://doi.org/10.1103/PhysRevLett.113.150402} {\bibfield  {journal}
  {\bibinfo  {journal} {Phys. Rev. Lett.}\ }\textbf {\bibinfo {volume} {113}},\
  \bibinfo {pages} {150402} (\bibinfo {year} {2014})}\BibitemShut {NoStop}%
\bibitem [{\citenamefont {{\'C}wikli{\'n}ski}\ \emph
  {et~al.}(2015)\citenamefont {{\'C}wikli{\'n}ski}, \citenamefont
  {Studzi{\'n}ski}, \citenamefont {Horodecki},\ and\ \citenamefont
  {Oppenheim}}]{Cwiklinski}%
  \BibitemOpen
  \bibfield  {author} {\bibinfo {author} {\bibfnamefont {P.}~\bibnamefont
  {{\'C}wikli{\'n}ski}}, \bibinfo {author} {\bibfnamefont {M.}~\bibnamefont
  {Studzi{\'n}ski}}, \bibinfo {author} {\bibfnamefont {M.}~\bibnamefont
  {Horodecki}},\ and\ \bibinfo {author} {\bibfnamefont {J.}~\bibnamefont
  {Oppenheim}},\ }\bibfield  {title} {\bibinfo {title} {Limitations on the
  evolution of quantum coherences: Towards fully quantum second laws of
  thermodynamics},\ }\href {https://doi.org/10.1103/PhysRevLett.115.210403}
  {\bibfield  {journal} {\bibinfo  {journal} {Phys. Rev. Lett.}\ }\textbf
  {\bibinfo {volume} {115}},\ \bibinfo {pages} {210403} (\bibinfo {year}
  {2015})}\BibitemShut {NoStop}%
\bibitem [{\citenamefont {Ficek}\ and\ \citenamefont {Swain}(2005)}]{Ficek}%
  \BibitemOpen
  \bibfield  {author} {\bibinfo {author} {\bibfnamefont {Z.}~\bibnamefont
  {Ficek}}\ and\ \bibinfo {author} {\bibfnamefont {S.}~\bibnamefont {Swain}},\
  }\href {https://www.springer.com/gp/book/9780387234654} {\emph {\bibinfo
  {title} {Quantum Interference and Coherence: Theory and Experiments}}},\
  Springer Series in Optical Sciences\ (\bibinfo  {publisher} {Springer},\
  \bibinfo {address} {New York},\ \bibinfo {year} {2005})\BibitemShut {NoStop}%
\bibitem [{\citenamefont {Baumgratz}\ \emph {et~al.}(2014)\citenamefont
  {Baumgratz}, \citenamefont {Cramer},\ and\ \citenamefont
  {Plenio}}]{Baumgratz}%
  \BibitemOpen
  \bibfield  {author} {\bibinfo {author} {\bibfnamefont {T.}~\bibnamefont
  {Baumgratz}}, \bibinfo {author} {\bibfnamefont {M.}~\bibnamefont {Cramer}},\
  and\ \bibinfo {author} {\bibfnamefont {M.~B.}\ \bibnamefont {Plenio}},\
  }\bibfield  {title} {\bibinfo {title} {Quantifying coherence},\ }\href
  {https://doi.org/10.1103/PhysRevLett.113.140401} {\bibfield  {journal}
  {\bibinfo  {journal} {Phys. Rev. Lett.}\ }\textbf {\bibinfo {volume} {113}},\
  \bibinfo {pages} {140401} (\bibinfo {year} {2014})}\BibitemShut {NoStop}%
\bibitem [{\citenamefont {Girolami}(2014)}]{Girolami}%
  \BibitemOpen
  \bibfield  {author} {\bibinfo {author} {\bibfnamefont {D.}~\bibnamefont
  {Girolami}},\ }\bibfield  {title} {\bibinfo {title} {Observable measure of
  quantum coherence in finite dimensional systems},\ }\href
  {https://doi.org/10.1103/PhysRevLett.113.170401} {\bibfield  {journal}
  {\bibinfo  {journal} {Phys. Rev. Lett.}\ }\textbf {\bibinfo {volume} {113}},\
  \bibinfo {pages} {170401} (\bibinfo {year} {2014})}\BibitemShut {NoStop}%
\bibitem [{\citenamefont {Streltsov}\ \emph {et~al.}(2015)\citenamefont
  {Streltsov}, \citenamefont {Singh}, \citenamefont {Dhar}, \citenamefont
  {Bera},\ and\ \citenamefont {Adesso}}]{Streltsov}%
  \BibitemOpen
  \bibfield  {author} {\bibinfo {author} {\bibfnamefont {A.}~\bibnamefont
  {Streltsov}}, \bibinfo {author} {\bibfnamefont {U.}~\bibnamefont {Singh}},
  \bibinfo {author} {\bibfnamefont {H.~S.}\ \bibnamefont {Dhar}}, \bibinfo
  {author} {\bibfnamefont {M.~N.}\ \bibnamefont {Bera}},\ and\ \bibinfo
  {author} {\bibfnamefont {G.}~\bibnamefont {Adesso}},\ }\bibfield  {title}
  {\bibinfo {title} {Measuring quantum coherence with entanglement},\ }\href
  {https://doi.org/10.1103/PhysRevLett.115.020403} {\bibfield  {journal}
  {\bibinfo  {journal} {Phys. Rev. Lett.}\ }\textbf {\bibinfo {volume} {115}},\
  \bibinfo {pages} {020403} (\bibinfo {year} {2015})}\BibitemShut {NoStop}%
\bibitem [{\citenamefont {Sachdev}(1999)}]{Sachdev}%
  \BibitemOpen
  \bibfield  {author} {\bibinfo {author} {\bibfnamefont {S.}~\bibnamefont
  {Sachdev}},\ }\href@noop {} {\emph {\bibinfo {title} {Quantum Phase
  Transitions}}}\ (\bibinfo  {publisher} {Cambridge University Press},\
  \bibinfo {address} {Cambridge},\ \bibinfo {year} {1999})\BibitemShut
  {NoStop}%
\bibitem [{\citenamefont {Chakrabarti}\ \emph {et~al.}(1996)\citenamefont
  {Chakrabarti}, \citenamefont {Dutta},\ and\ \citenamefont
  {Sen}}]{Chakrabarti}%
  \BibitemOpen
  \bibfield  {author} {\bibinfo {author} {\bibfnamefont {B.~K.}\ \bibnamefont
  {Chakrabarti}}, \bibinfo {author} {\bibfnamefont {A.}~\bibnamefont {Dutta}},\
  and\ \bibinfo {author} {\bibfnamefont {P.}~\bibnamefont {Sen}},\ }\href@noop
  {} {\emph {\bibinfo {title} {Quantum Ising Phases and Transitions in
  Transverse Ising Models}}}\ (\bibinfo  {publisher} {Springer},\ \bibinfo
  {address} {Heidelberg},\ \bibinfo {year} {1996})\BibitemShut {NoStop}%
\bibitem [{\citenamefont {Continentino}(2001)}]{Continentino}%
  \BibitemOpen
  \bibfield  {author} {\bibinfo {author} {\bibfnamefont {M.~A.}\ \bibnamefont
  {Continentino}},\ }\href@noop {} {\emph {\bibinfo {title} {Quantum Scaling in
  Many-Body Systems}}}\ (\bibinfo  {publisher} {World Scientific},\ \bibinfo
  {address} {Singapore},\ \bibinfo {year} {2001})\BibitemShut {NoStop}%
\bibitem [{\citenamefont {Osterloh}\ \emph {et~al.}(2002)\citenamefont
  {Osterloh}, \citenamefont {Amico}, \citenamefont {Falci},\ and\ \citenamefont
  {Fazio}}]{Osterloh2002}%
  \BibitemOpen
  \bibfield  {author} {\bibinfo {author} {\bibfnamefont {A.}~\bibnamefont
  {Osterloh}}, \bibinfo {author} {\bibfnamefont {L.}~\bibnamefont {Amico}},
  \bibinfo {author} {\bibfnamefont {G.}~\bibnamefont {Falci}},\ and\ \bibinfo
  {author} {\bibfnamefont {R.}~\bibnamefont {Fazio}},\ }\bibfield  {title}
  {\bibinfo {title} {Scaling of entanglement close to a quantum phase
  transition},\ }\href {https://doi.org/doi.org/10.1038/416608a} {\bibfield
  {journal} {\bibinfo  {journal} {Nature}\ }\textbf {\bibinfo {volume} {416}},\
  \bibinfo {pages} {608} (\bibinfo {year} {2002})}\BibitemShut {NoStop}%
\bibitem [{\citenamefont {Karpat}\ \emph {et~al.}(2014)\citenamefont {Karpat},
  \citenamefont {\c{C}akmak},\ and\ \citenamefont {Fanchini}}]{Karpat}%
  \BibitemOpen
  \bibfield  {author} {\bibinfo {author} {\bibfnamefont {G.}~\bibnamefont
  {Karpat}}, \bibinfo {author} {\bibfnamefont {B.}~\bibnamefont {\c{C}akmak}},\
  and\ \bibinfo {author} {\bibfnamefont {F.~F.}\ \bibnamefont {Fanchini}},\
  }\bibfield  {title} {\bibinfo {title} {Quantum coherence and uncertainty in
  the anisotropic $\mathrm{XY}$ chain},\ }\href
  {https://doi.org/10.1103/PhysRevB.90.104431} {\bibfield  {journal} {\bibinfo
  {journal} {Phys. Rev. B}\ }\textbf {\bibinfo {volume} {90}},\ \bibinfo
  {pages} {104431} (\bibinfo {year} {2014})}\BibitemShut {NoStop}%
\bibitem [{\citenamefont {Invernizzi}\ \emph {et~al.}(2008)\citenamefont
  {Invernizzi}, \citenamefont {Korbman}, \citenamefont {Campos~Venuti},\ and\
  \citenamefont {Paris}}]{Invernizzi}%
  \BibitemOpen
  \bibfield  {author} {\bibinfo {author} {\bibfnamefont {C.}~\bibnamefont
  {Invernizzi}}, \bibinfo {author} {\bibfnamefont {M.}~\bibnamefont {Korbman}},
  \bibinfo {author} {\bibfnamefont {L.}~\bibnamefont {Campos~Venuti}},\ and\
  \bibinfo {author} {\bibfnamefont {M.~G.~A.}\ \bibnamefont {Paris}},\
  }\bibfield  {title} {\bibinfo {title} {Optimal quantum estimation in spin
  systems at criticality},\ }\href {https://doi.org/10.1103/PhysRevA.78.042106}
  {\bibfield  {journal} {\bibinfo  {journal} {Phys. Rev. A}\ }\textbf {\bibinfo
  {volume} {78}},\ \bibinfo {pages} {042106} (\bibinfo {year}
  {2008})}\BibitemShut {NoStop}%
\bibitem [{\citenamefont {Sun}\ \emph {et~al.}(2010)\citenamefont {Sun},
  \citenamefont {Ma}, \citenamefont {Lu},\ and\ \citenamefont {Wang}}]{Sun}%
  \BibitemOpen
  \bibfield  {author} {\bibinfo {author} {\bibfnamefont {Z.}~\bibnamefont
  {Sun}}, \bibinfo {author} {\bibfnamefont {J.}~\bibnamefont {Ma}}, \bibinfo
  {author} {\bibfnamefont {X.-M.}\ \bibnamefont {Lu}},\ and\ \bibinfo {author}
  {\bibfnamefont {X.}~\bibnamefont {Wang}},\ }\bibfield  {title} {\bibinfo
  {title} {Fisher information in a quantum-critical environment},\ }\href
  {https://doi.org/10.1103/PhysRevA.82.022306} {\bibfield  {journal} {\bibinfo
  {journal} {Phys. Rev. A}\ }\textbf {\bibinfo {volume} {82}},\ \bibinfo
  {pages} {022306} (\bibinfo {year} {2010})}\BibitemShut {NoStop}%
\bibitem [{\citenamefont {Xi}\ \emph {et~al.}(2015)\citenamefont {Xi},
  \citenamefont {Li},\ and\ \citenamefont {Fan}}]{Xi2015}%
  \BibitemOpen
  \bibfield  {author} {\bibinfo {author} {\bibfnamefont {Z.}~\bibnamefont
  {Xi}}, \bibinfo {author} {\bibfnamefont {Y.}~\bibnamefont {Li}},\ and\
  \bibinfo {author} {\bibfnamefont {H.}~\bibnamefont {Fan}},\ }\bibfield
  {title} {\bibinfo {title} {Quantum coherence and correlations in quantum
  system},\ }\href {https://doi.org/10.1038/srep10922} {\bibfield  {journal}
  {\bibinfo  {journal} {Scientific Reports}\ }\textbf {\bibinfo {volume} {5}},\
  \bibinfo {pages} {10922} (\bibinfo {year} {2015})}\BibitemShut {NoStop}%
\bibitem [{\citenamefont {Hu}\ and\ \citenamefont {Fan}(2016)}]{Hu2016}%
  \BibitemOpen
  \bibfield  {author} {\bibinfo {author} {\bibfnamefont {M.-L.}\ \bibnamefont
  {Hu}}\ and\ \bibinfo {author} {\bibfnamefont {H.}~\bibnamefont {Fan}},\
  }\bibfield  {title} {\bibinfo {title} {Evolution equation for quantum
  coherence},\ }\href {https://doi.org/10.1038/srep29260} {\bibfield  {journal}
  {\bibinfo  {journal} {Scientific Reports}\ }\textbf {\bibinfo {volume} {6}},\
  \bibinfo {pages} {29260} (\bibinfo {year} {2016})}\BibitemShut {NoStop}%
\bibitem [{\citenamefont {Rana}\ \emph {et~al.}(2016)\citenamefont {Rana},
  \citenamefont {Parashar},\ and\ \citenamefont {Lewenstein}}]{Rana2016}%
  \BibitemOpen
  \bibfield  {author} {\bibinfo {author} {\bibfnamefont {S.}~\bibnamefont
  {Rana}}, \bibinfo {author} {\bibfnamefont {P.}~\bibnamefont {Parashar}},\
  and\ \bibinfo {author} {\bibfnamefont {M.}~\bibnamefont {Lewenstein}},\
  }\bibfield  {title} {\bibinfo {title} {Trace-distance measure of coherence},\
  }\href {https://doi.org/10.1103/PhysRevA.93.012110} {\bibfield  {journal}
  {\bibinfo  {journal} {Phys. Rev. A}\ }\textbf {\bibinfo {volume} {93}},\
  \bibinfo {pages} {012110} (\bibinfo {year} {2016})}\BibitemShut {NoStop}%
\bibitem [{\citenamefont {Chen}\ \emph {et~al.}(2016)\citenamefont {Chen},
  \citenamefont {Cui}, \citenamefont {Zhang},\ and\ \citenamefont
  {Fan}}]{chen}%
  \BibitemOpen
  \bibfield  {author} {\bibinfo {author} {\bibfnamefont {J.-J.}\ \bibnamefont
  {Chen}}, \bibinfo {author} {\bibfnamefont {J.}~\bibnamefont {Cui}}, \bibinfo
  {author} {\bibfnamefont {Y.-R.}\ \bibnamefont {Zhang}},\ and\ \bibinfo
  {author} {\bibfnamefont {H.}~\bibnamefont {Fan}},\ }\bibfield  {title}
  {\bibinfo {title} {Coherence susceptibility as a probe of quantum phase
  transitions},\ }\href {https://doi.org/10.1103/PhysRevA.94.022112} {\bibfield
   {journal} {\bibinfo  {journal} {Phys. Rev. A}\ }\textbf {\bibinfo {volume}
  {94}},\ \bibinfo {pages} {022112} (\bibinfo {year} {2016})}\BibitemShut
  {NoStop}%
\bibitem [{\citenamefont {Qin}\ \emph {et~al.}(2018)\citenamefont {Qin},
  \citenamefont {Ren},\ and\ \citenamefont {Zhang}}]{Qin2018}%
  \BibitemOpen
  \bibfield  {author} {\bibinfo {author} {\bibfnamefont {M.}~\bibnamefont
  {Qin}}, \bibinfo {author} {\bibfnamefont {Z.}~\bibnamefont {Ren}},\ and\
  \bibinfo {author} {\bibfnamefont {X.}~\bibnamefont {Zhang}},\ }\bibfield
  {title} {\bibinfo {title} {Dynamics of quantum coherence and quantum phase
  transitions in $\mathrm{XY}$ spin systems},\ }\href
  {https://doi.org/10.1103/PhysRevA.98.012303} {\bibfield  {journal} {\bibinfo
  {journal} {Phys. Rev. A}\ }\textbf {\bibinfo {volume} {98}},\ \bibinfo
  {pages} {012303} (\bibinfo {year} {2018})}\BibitemShut {NoStop}%
\bibitem [{\citenamefont {Gu}\ \emph {et~al.}(2003)\citenamefont {Gu},
  \citenamefont {Lin},\ and\ \citenamefont {Li}}]{Gu2003}%
  \BibitemOpen
  \bibfield  {author} {\bibinfo {author} {\bibfnamefont {S.-J.}\ \bibnamefont
  {Gu}}, \bibinfo {author} {\bibfnamefont {H.-Q.}\ \bibnamefont {Lin}},\ and\
  \bibinfo {author} {\bibfnamefont {Y.-Q.}\ \bibnamefont {Li}},\ }\bibfield
  {title} {\bibinfo {title} {Entanglement, quantum phase transition, and
  scaling in the $\mathrm{XXZ}$ chain},\ }\href
  {https://doi.org/10.1103/PhysRevA.68.042330} {\bibfield  {journal} {\bibinfo
  {journal} {Phys. Rev. A}\ }\textbf {\bibinfo {volume} {68}},\ \bibinfo
  {pages} {042330} (\bibinfo {year} {2003})}\BibitemShut {NoStop}%
\bibitem [{\citenamefont {Mondal}\ \emph {et~al.}(2017)\citenamefont {Mondal},
  \citenamefont {Pramanik},\ and\ \citenamefont {Pati}}]{Mondal}%
  \BibitemOpen
  \bibfield  {author} {\bibinfo {author} {\bibfnamefont {D.}~\bibnamefont
  {Mondal}}, \bibinfo {author} {\bibfnamefont {T.}~\bibnamefont {Pramanik}},\
  and\ \bibinfo {author} {\bibfnamefont {A.~K.}\ \bibnamefont {Pati}},\
  }\bibfield  {title} {\bibinfo {title} {Nonlocal advantage of quantum
  coherence},\ }\href {https://doi.org/10.1103/PhysRevA.95.010301} {\bibfield
  {journal} {\bibinfo  {journal} {Phys. Rev. A}\ }\textbf {\bibinfo {volume}
  {95}},\ \bibinfo {pages} {010301} (\bibinfo {year} {2017})}\BibitemShut
  {NoStop}%
\bibitem [{\citenamefont {Hu}\ \emph {et~al.}(2020)\citenamefont {Hu},
  \citenamefont {Gao},\ and\ \citenamefont {Fan}}]{Hu2020}%
  \BibitemOpen
  \bibfield  {author} {\bibinfo {author} {\bibfnamefont {M.-L.}\ \bibnamefont
  {Hu}}, \bibinfo {author} {\bibfnamefont {Y.-Y.}\ \bibnamefont {Gao}},\ and\
  \bibinfo {author} {\bibfnamefont {H.}~\bibnamefont {Fan}},\ }\bibfield
  {title} {\bibinfo {title} {Steered quantum coherence as a signature of
  quantum phase transitions in spin chains},\ }\href
  {https://doi.org/10.1103/PhysRevA.101.032305} {\bibfield  {journal} {\bibinfo
   {journal} {Phys. Rev. A}\ }\textbf {\bibinfo {volume} {101}},\ \bibinfo
  {pages} {032305} (\bibinfo {year} {2020})}\BibitemShut {NoStop}%
\bibitem [{\citenamefont {Dai}\ \emph {et~al.}(2022)\citenamefont {Dai},
  \citenamefont {Fu},\ and\ \citenamefont {Luo}}]{Dai2022}%
  \BibitemOpen
  \bibfield  {author} {\bibinfo {author} {\bibfnamefont {H.}~\bibnamefont
  {Dai}}, \bibinfo {author} {\bibfnamefont {S.}~\bibnamefont {Fu}},\ and\
  \bibinfo {author} {\bibfnamefont {S.}~\bibnamefont {Luo}},\ }\bibfield
  {title} {\bibinfo {title} {Detecting magic states via characteristic
  functions},\ }\href {https://doi.org/10.1007/s10773-022-05027-8} {\bibfield
  {journal} {\bibinfo  {journal} {International Journal of Theoretical
  Physics}\ }\textbf {\bibinfo {volume} {61}},\ \bibinfo {pages} {35} (\bibinfo
  {year} {2022})}\BibitemShut {NoStop}%
\bibitem [{\citenamefont {Shor}(1996)}]{Shor}%
  \BibitemOpen
  \bibfield  {author} {\bibinfo {author} {\bibfnamefont {P.}~\bibnamefont
  {Shor}},\ }\bibfield  {title} {\bibinfo {title} {Fault-tolerant quantum
  computation},\ }\bibfield  {journal} {\bibinfo  {journal}
  {arXiv:quant-ph/9605011v2}\ }\href
  {https://doi.org/https://doi.org/10.48550/arXiv.quant-ph/9605011}
  {https://doi.org/10.48550/arXiv.quant-ph/9605011} (\bibinfo {year}
  {1996})\BibitemShut {NoStop}%
\bibitem [{\citenamefont {Preskill}(1997)}]{Preskill1997}%
  \BibitemOpen
  \bibfield  {author} {\bibinfo {author} {\bibfnamefont {J.}~\bibnamefont
  {Preskill}},\ }\bibfield  {title} {\bibinfo {title} {Fault-tolerant quantum
  computation},\ }\bibfield  {journal} {\bibinfo  {journal}
  {arXiv:quant-ph/9712048}\ }\href
  {https://doi.org/https://doi.org/10.48550/arXiv.quant-ph/9712048}
  {https://doi.org/10.48550/arXiv.quant-ph/9712048} (\bibinfo {year}
  {1997})\BibitemShut {NoStop}%
\bibitem [{\citenamefont {Knill}(2005)}]{Knill}%
  \BibitemOpen
  \bibfield  {author} {\bibinfo {author} {\bibfnamefont {E.}~\bibnamefont
  {Knill}},\ }\bibfield  {title} {\bibinfo {title} {Quantum computing with
  realistically noisy devices},\ }\href {https://doi.org/10.1038/nature03350}
  {\bibfield  {journal} {\bibinfo  {journal} {Nature}\ }\textbf {\bibinfo
  {volume} {434}},\ \bibinfo {pages} {39} (\bibinfo {year} {2005})}\BibitemShut
  {NoStop}%
\bibitem [{\citenamefont {Bravyi}\ and\ \citenamefont
  {Kitaev}(2005)}]{Bravyi2005}%
  \BibitemOpen
  \bibfield  {author} {\bibinfo {author} {\bibfnamefont {S.}~\bibnamefont
  {Bravyi}}\ and\ \bibinfo {author} {\bibfnamefont {A.}~\bibnamefont
  {Kitaev}},\ }\bibfield  {title} {\bibinfo {title} {Universal quantum
  computation with ideal clifford gates and noisy ancillas},\ }\href
  {https://doi.org/10.1103/PhysRevA.71.022316} {\bibfield  {journal} {\bibinfo
  {journal} {Phys. Rev. A}\ }\textbf {\bibinfo {volume} {71}},\ \bibinfo
  {pages} {022316} (\bibinfo {year} {2005})}\BibitemShut {NoStop}%
\bibitem [{\citenamefont {Campbell}\ \emph {et~al.}(2017)\citenamefont
  {Campbell}, \citenamefont {Terhal},\ and\ \citenamefont
  {Vuillot}}]{Campbell2017}%
  \BibitemOpen
  \bibfield  {author} {\bibinfo {author} {\bibfnamefont {E.}~\bibnamefont
  {Campbell}}, \bibinfo {author} {\bibfnamefont {B.}~\bibnamefont {Terhal}},\
  and\ \bibinfo {author} {\bibfnamefont {C.}~\bibnamefont {Vuillot}},\
  }\bibfield  {title} {\bibinfo {title} {Roads towards fault-tolerant universal
  quantum computation},\ }\href {https://doi.org/10.1038/nature23460}
  {\bibfield  {journal} {\bibinfo  {journal} {Nature}\ }\textbf {\bibinfo
  {volume} {549}},\ \bibinfo {pages} {172} (\bibinfo {year}
  {2017})}\BibitemShut {NoStop}%
\bibitem [{\citenamefont {Gottesman}(1998)}]{Gottesman1998}%
  \BibitemOpen
  \bibfield  {author} {\bibinfo {author} {\bibfnamefont {D.}~\bibnamefont
  {Gottesman}},\ }\bibfield  {title} {\bibinfo {title} {The heisenberg
  representation of quantum computers},\ }\bibfield  {journal} {\bibinfo
  {journal} {arXiv:quant-ph/9807006}\ }\href
  {https://doi.org/https://doi.org/10.48550/arXiv.quant-ph/9807006}
  {https://doi.org/10.48550/arXiv.quant-ph/9807006} (\bibinfo {year}
  {1998})\BibitemShut {NoStop}%
\bibitem [{\citenamefont {Veitch}\ \emph {et~al.}(2013)\citenamefont {Veitch},
  \citenamefont {Mousavian}, \citenamefont {Gottesman},\ and\ \citenamefont
  {Emerson}}]{Veitch2013}%
  \BibitemOpen
  \bibfield  {author} {\bibinfo {author} {\bibfnamefont {V.}~\bibnamefont
  {Veitch}}, \bibinfo {author} {\bibfnamefont {S.~A.}\ \bibnamefont
  {Mousavian}}, \bibinfo {author} {\bibfnamefont {D.}~\bibnamefont
  {Gottesman}},\ and\ \bibinfo {author} {\bibfnamefont {J.}~\bibnamefont
  {Emerson}},\ }\bibfield  {title} {\bibinfo {title} {The resource theory of
  stabilizer computation},\ }\href
  {https://doi.org/10.1088/1367-2630/16/1/013009} {\bibfield  {journal}
  {\bibinfo  {journal} {New J. Phys.}\ }\textbf {\bibinfo {volume} {16}},\
  \bibinfo {pages} {013009} (\bibinfo {year} {2013})}\BibitemShut {NoStop}%
\bibitem [{\citenamefont {Bravyi}\ \emph {et~al.}(2016)\citenamefont {Bravyi},
  \citenamefont {Smith},\ and\ \citenamefont {Smolin}}]{Bravyi2016}%
  \BibitemOpen
  \bibfield  {author} {\bibinfo {author} {\bibfnamefont {S.}~\bibnamefont
  {Bravyi}}, \bibinfo {author} {\bibfnamefont {G.}~\bibnamefont {Smith}},\ and\
  \bibinfo {author} {\bibfnamefont {J.~A.}\ \bibnamefont {Smolin}},\ }\bibfield
   {title} {\bibinfo {title} {Trading classical and quantum resources},\ }\href
  {https://doi.org/10.1103/PhysRevX.6.021043} {\bibfield  {journal} {\bibinfo
  {journal} {Phys. Rev. X}\ }\textbf {\bibinfo {volume} {6}},\ \bibinfo {pages}
  {021043} (\bibinfo {year} {2016})}\BibitemShut {NoStop}%
\bibitem [{\citenamefont {Howard}\ and\ \citenamefont
  {Campbell}(2017)}]{Howard2017}%
  \BibitemOpen
  \bibfield  {author} {\bibinfo {author} {\bibfnamefont {M.}~\bibnamefont
  {Howard}}\ and\ \bibinfo {author} {\bibfnamefont {E.}~\bibnamefont
  {Campbell}},\ }\bibfield  {title} {\bibinfo {title} {Application of a
  resource theory for magic states to fault-tolerant quantum computing},\
  }\href {https://doi.org/10.1103/PhysRevLett.118.090501} {\bibfield  {journal}
  {\bibinfo  {journal} {Phys. Rev. Lett.}\ }\textbf {\bibinfo {volume} {118}},\
  \bibinfo {pages} {090501} (\bibinfo {year} {2017})}\BibitemShut {NoStop}%
\bibitem [{\citenamefont {Ahmadi}\ \emph {et~al.}(2018)\citenamefont {Ahmadi},
  \citenamefont {Dang}, \citenamefont {Gour},\ and\ \citenamefont
  {Sanders}}]{Ahmadi2018}%
  \BibitemOpen
  \bibfield  {author} {\bibinfo {author} {\bibfnamefont {M.}~\bibnamefont
  {Ahmadi}}, \bibinfo {author} {\bibfnamefont {H.~B.}\ \bibnamefont {Dang}},
  \bibinfo {author} {\bibfnamefont {G.}~\bibnamefont {Gour}},\ and\ \bibinfo
  {author} {\bibfnamefont {B.~C.}\ \bibnamefont {Sanders}},\ }\bibfield
  {title} {\bibinfo {title} {Quantification and manipulation of magic states},\
  }\href {https://doi.org/10.1103/PhysRevA.97.062332} {\bibfield  {journal}
  {\bibinfo  {journal} {Phys. Rev. A}\ }\textbf {\bibinfo {volume} {97}},\
  \bibinfo {pages} {062332} (\bibinfo {year} {2018})}\BibitemShut {NoStop}%
\bibitem [{\citenamefont {Heinrich}\ and\ \citenamefont
  {Gross}(2019)}]{Heinrich2019}%
  \BibitemOpen
  \bibfield  {author} {\bibinfo {author} {\bibfnamefont {M.}~\bibnamefont
  {Heinrich}}\ and\ \bibinfo {author} {\bibfnamefont {D.}~\bibnamefont
  {Gross}},\ }\bibfield  {title} {\bibinfo {title} {Robustness of {M}agic and
  {S}ymmetries of the {S}tabiliser {P}olytope},\ }\href
  {https://doi.org/10.22331/q-2019-04-08-132} {\bibfield  {journal} {\bibinfo
  {journal} {{Quantum}}\ }\textbf {\bibinfo {volume} {3}},\ \bibinfo {pages}
  {132} (\bibinfo {year} {2019})}\BibitemShut {NoStop}%
\bibitem [{\citenamefont {Seddon}\ and\ \citenamefont
  {Campbell}(2019)}]{Seddon2019}%
  \BibitemOpen
  \bibfield  {author} {\bibinfo {author} {\bibfnamefont {J.~R.}\ \bibnamefont
  {Seddon}}\ and\ \bibinfo {author} {\bibfnamefont {E.~T.}\ \bibnamefont
  {Campbell}},\ }\bibfield  {title} {\bibinfo {title} {Quantifying magic for
  multi-qubit operations},\ }\href {https://doi.org/10.1098/rspa.2019.0251}
  {\bibfield  {journal} {\bibinfo  {journal} {Proceedings of the Royal Society
  A: Mathematical, Physical and Engineering Sciences}\ }\textbf {\bibinfo
  {volume} {475}},\ \bibinfo {pages} {20190251} (\bibinfo {year}
  {2019})}\BibitemShut {NoStop}%
\bibitem [{\citenamefont {Bravyi}\ \emph {et~al.}(2019)\citenamefont {Bravyi},
  \citenamefont {Browne}, \citenamefont {Calpin}, \citenamefont {Campbell},
  \citenamefont {Gosset},\ and\ \citenamefont {Howard}}]{Bravyi2019}%
  \BibitemOpen
  \bibfield  {author} {\bibinfo {author} {\bibfnamefont {S.}~\bibnamefont
  {Bravyi}}, \bibinfo {author} {\bibfnamefont {D.}~\bibnamefont {Browne}},
  \bibinfo {author} {\bibfnamefont {P.}~\bibnamefont {Calpin}}, \bibinfo
  {author} {\bibfnamefont {E.}~\bibnamefont {Campbell}}, \bibinfo {author}
  {\bibfnamefont {D.}~\bibnamefont {Gosset}},\ and\ \bibinfo {author}
  {\bibfnamefont {M.}~\bibnamefont {Howard}},\ }\bibfield  {title} {\bibinfo
  {title} {Simulation of quantum circuits by low-rank stabilizer
  decompositions},\ }\href {https://doi.org/10.22331/q-2019-09-02-181}
  {\bibfield  {journal} {\bibinfo  {journal} {{Quantum}}\ }\textbf {\bibinfo
  {volume} {3}},\ \bibinfo {pages} {181} (\bibinfo {year} {2019})}\BibitemShut
  {NoStop}%
\bibitem [{\citenamefont {Wang}\ \emph {et~al.}(2020)\citenamefont {Wang},
  \citenamefont {Wilde},\ and\ \citenamefont {Su}}]{Wang2020}%
  \BibitemOpen
  \bibfield  {author} {\bibinfo {author} {\bibfnamefont {X.}~\bibnamefont
  {Wang}}, \bibinfo {author} {\bibfnamefont {M.~M.}\ \bibnamefont {Wilde}},\
  and\ \bibinfo {author} {\bibfnamefont {Y.}~\bibnamefont {Su}},\ }\bibfield
  {title} {\bibinfo {title} {Efficiently computable bounds for magic state
  distillation},\ }\href {https://doi.org/10.1103/PhysRevLett.124.090505}
  {\bibfield  {journal} {\bibinfo  {journal} {Phys. Rev. Lett.}\ }\textbf
  {\bibinfo {volume} {124}},\ \bibinfo {pages} {090505} (\bibinfo {year}
  {2020})}\BibitemShut {NoStop}%
\bibitem [{\citenamefont {Seddon}\ \emph {et~al.}(2021)\citenamefont {Seddon},
  \citenamefont {Regula}, \citenamefont {Pashayan}, \citenamefont {Ouyang},\
  and\ \citenamefont {Campbell}}]{Seddon2021}%
  \BibitemOpen
  \bibfield  {author} {\bibinfo {author} {\bibfnamefont {J.~R.}\ \bibnamefont
  {Seddon}}, \bibinfo {author} {\bibfnamefont {B.}~\bibnamefont {Regula}},
  \bibinfo {author} {\bibfnamefont {H.}~\bibnamefont {Pashayan}}, \bibinfo
  {author} {\bibfnamefont {Y.}~\bibnamefont {Ouyang}},\ and\ \bibinfo {author}
  {\bibfnamefont {E.~T.}\ \bibnamefont {Campbell}},\ }\bibfield  {title}
  {\bibinfo {title} {Quantifying quantum speedups: Improved classical
  simulation from tighter magic monotones},\ }\href
  {https://doi.org/10.1103/PRXQuantum.2.010345} {\bibfield  {journal} {\bibinfo
   {journal} {PRX Quantum}\ }\textbf {\bibinfo {volume} {2}},\ \bibinfo {pages}
  {010345} (\bibinfo {year} {2021})}\BibitemShut {NoStop}%
\bibitem [{\citenamefont {Heimendahl}\ \emph {et~al.}(2021)\citenamefont
  {Heimendahl}, \citenamefont {Montealegre-Mora}, \citenamefont {Vallentin},\
  and\ \citenamefont {Gross}}]{Heimendahl2021}%
  \BibitemOpen
  \bibfield  {author} {\bibinfo {author} {\bibfnamefont {A.}~\bibnamefont
  {Heimendahl}}, \bibinfo {author} {\bibfnamefont {F.}~\bibnamefont
  {Montealegre-Mora}}, \bibinfo {author} {\bibfnamefont {F.}~\bibnamefont
  {Vallentin}},\ and\ \bibinfo {author} {\bibfnamefont {D.}~\bibnamefont
  {Gross}},\ }\bibfield  {title} {\bibinfo {title} {Stabilizer extent is not
  multiplicative},\ }\href {https://doi.org/10.22331/q-2021-02-24-400}
  {\bibfield  {journal} {\bibinfo  {journal} {{Quantum}}\ }\textbf {\bibinfo
  {volume} {5}},\ \bibinfo {pages} {400} (\bibinfo {year} {2021})}\BibitemShut
  {NoStop}%
\bibitem [{\citenamefont {Liu}\ and\ \citenamefont {Winter}(2022)}]{Liu2022}%
  \BibitemOpen
  \bibfield  {author} {\bibinfo {author} {\bibfnamefont {Z.-W.}\ \bibnamefont
  {Liu}}\ and\ \bibinfo {author} {\bibfnamefont {A.}~\bibnamefont {Winter}},\
  }\bibfield  {title} {\bibinfo {title} {Many-body quantum magic},\ }\href
  {https://doi.org/10.1103/PRXQuantum.3.020333} {\bibfield  {journal} {\bibinfo
   {journal} {PRX Quantum}\ }\textbf {\bibinfo {volume} {3}},\ \bibinfo {pages}
  {020333} (\bibinfo {year} {2022})}\BibitemShut {NoStop}%
\bibitem [{\citenamefont {Fu}\ \emph {et~al.}(2022)\citenamefont {Fu},
  \citenamefont {Li},\ and\ \citenamefont {Luo}}]{Fu2022}%
  \BibitemOpen
  \bibfield  {author} {\bibinfo {author} {\bibfnamefont {S.}~\bibnamefont
  {Fu}}, \bibinfo {author} {\bibfnamefont {X.}~\bibnamefont {Li}},\ and\
  \bibinfo {author} {\bibfnamefont {S.}~\bibnamefont {Luo}},\ }\bibfield
  {title} {\bibinfo {title} {Detecting quantum phase transition via magic
  resource in the $\mathrm{XY}$ spin model},\ }\href
  {https://doi.org/10.1103/PhysRevA.106.062405} {\bibfield  {journal} {\bibinfo
   {journal} {Phys. Rev. A}\ }\textbf {\bibinfo {volume} {106}},\ \bibinfo
  {pages} {062405} (\bibinfo {year} {2022})}\BibitemShut {NoStop}%
\bibitem [{\citenamefont {Mishra}\ \emph {et~al.}(2016)\citenamefont {Mishra},
  \citenamefont {Rakshit},\ and\ \citenamefont {Prabhu}}]{Mishra2016}%
  \BibitemOpen
  \bibfield  {author} {\bibinfo {author} {\bibfnamefont {U.}~\bibnamefont
  {Mishra}}, \bibinfo {author} {\bibfnamefont {D.}~\bibnamefont {Rakshit}},\
  and\ \bibinfo {author} {\bibfnamefont {R.}~\bibnamefont {Prabhu}},\
  }\bibfield  {title} {\bibinfo {title} {Survival of time-evolved quantum
  correlations depending on whether quenching is across a critical point in an
  $\mathrm{XY}$ spin chain},\ }\href
  {https://doi.org/10.1103/PhysRevA.93.042322} {\bibfield  {journal} {\bibinfo
  {journal} {Phys. Rev. A}\ }\textbf {\bibinfo {volume} {93}},\ \bibinfo
  {pages} {042322} (\bibinfo {year} {2016})}\BibitemShut {NoStop}%
\bibitem [{\citenamefont {Jafari}(2010)}]{Jafari:2010aa}%
  \BibitemOpen
  \bibfield  {author} {\bibinfo {author} {\bibfnamefont {R.}~\bibnamefont
  {Jafari}},\ }\bibfield  {title} {\bibinfo {title} {Low-energy-state dynamics
  of entanglement for spin systems},\ }\href
  {https://doi.org/10.1103/PhysRevA.82.052317} {\bibfield  {journal} {\bibinfo
  {journal} {Phys. Rev. A}\ }\textbf {\bibinfo {volume} {82}},\ \bibinfo
  {pages} {052317} (\bibinfo {year} {2010})}\BibitemShut {NoStop}%
\bibitem [{\citenamefont {Jafari}\ and\ \citenamefont
  {Akbari}(2020)}]{Jafari:2020aa}%
  \BibitemOpen
  \bibfield  {author} {\bibinfo {author} {\bibfnamefont {R.}~\bibnamefont
  {Jafari}}\ and\ \bibinfo {author} {\bibfnamefont {A.}~\bibnamefont
  {Akbari}},\ }\bibfield  {title} {\bibinfo {title} {Dynamics of quantum
  coherence and quantum fisher information after a sudden quench},\ }\href
  {https://doi.org/10.1103/PhysRevA.101.062105} {\bibfield  {journal} {\bibinfo
   {journal} {Phys. Rev. A}\ }\textbf {\bibinfo {volume} {101}},\ \bibinfo
  {pages} {062105} (\bibinfo {year} {2020})}\BibitemShut {NoStop}%
\bibitem [{\citenamefont {Jafari}\ \emph {et~al.}(2019)\citenamefont {Jafari},
  \citenamefont {Johannesson}, \citenamefont {Langari},\ and\ \citenamefont
  {Martin-Delgado}}]{Jafari:2019aa}%
  \BibitemOpen
  \bibfield  {author} {\bibinfo {author} {\bibfnamefont {R.}~\bibnamefont
  {Jafari}}, \bibinfo {author} {\bibfnamefont {H.}~\bibnamefont {Johannesson}},
  \bibinfo {author} {\bibfnamefont {A.}~\bibnamefont {Langari}},\ and\ \bibinfo
  {author} {\bibfnamefont {M.~A.}\ \bibnamefont {Martin-Delgado}},\ }\bibfield
  {title} {\bibinfo {title} {Quench dynamics and zero-energy modes: The case of
  the creutz model},\ }\href {https://doi.org/10.1103/PhysRevB.99.054302}
  {\bibfield  {journal} {\bibinfo  {journal} {Phys. Rev. B}\ }\textbf {\bibinfo
  {volume} {99}},\ \bibinfo {pages} {054302} (\bibinfo {year}
  {2019})}\BibitemShut {NoStop}%
\bibitem [{\citenamefont {Mishra}\ \emph {et~al.}(2018)\citenamefont {Mishra},
  \citenamefont {Cheraghi}, \citenamefont {Mahdavifar}, \citenamefont
  {Jafari},\ and\ \citenamefont {Akbari}}]{Mishra:2018aa}%
  \BibitemOpen
  \bibfield  {author} {\bibinfo {author} {\bibfnamefont {U.}~\bibnamefont
  {Mishra}}, \bibinfo {author} {\bibfnamefont {H.}~\bibnamefont {Cheraghi}},
  \bibinfo {author} {\bibfnamefont {S.}~\bibnamefont {Mahdavifar}}, \bibinfo
  {author} {\bibfnamefont {R.}~\bibnamefont {Jafari}},\ and\ \bibinfo {author}
  {\bibfnamefont {A.}~\bibnamefont {Akbari}},\ }\bibfield  {title} {\bibinfo
  {title} {Dynamical quantum correlations after sudden quenches},\ }\href
  {https://doi.org/10.1103/PhysRevA.98.052338} {\bibfield  {journal} {\bibinfo
  {journal} {Phys. Rev. A}\ }\textbf {\bibinfo {volume} {98}},\ \bibinfo
  {pages} {052338} (\bibinfo {year} {2018})}\BibitemShut {NoStop}%
\bibitem [{\citenamefont {McCoy}(1968)}]{McCoy:1968aa}%
  \BibitemOpen
  \bibfield  {author} {\bibinfo {author} {\bibfnamefont {B.~M.}\ \bibnamefont
  {McCoy}},\ }\bibfield  {title} {\bibinfo {title} {Spin correlation functions
  of the $\mathrm{X-Y}$ model},\ }\href
  {https://doi.org/10.1103/PhysRev.173.531} {\bibfield  {journal} {\bibinfo
  {journal} {Phys. Rev.}\ }\textbf {\bibinfo {volume} {173}},\ \bibinfo {pages}
  {531} (\bibinfo {year} {1968})}\BibitemShut {NoStop}%
\bibitem [{\citenamefont {Pfeuty}(1970)}]{phase}%
  \BibitemOpen
  \bibfield  {author} {\bibinfo {author} {\bibfnamefont {P.}~\bibnamefont
  {Pfeuty}},\ }\bibfield  {title} {\bibinfo {title} {The one-dimensional ising
  model with a transverse field},\ }\href
  {https://doi.org/10.1016/0003-4916(70)90270-8} {\bibfield  {journal}
  {\bibinfo  {journal} {Annals of Physics}\ }\textbf {\bibinfo {volume} {57}},\
  \bibinfo {pages} {79} (\bibinfo {year} {1970})}\BibitemShut {NoStop}%
\bibitem [{\citenamefont {Sadiek}\ \emph {et~al.}(2010)\citenamefont {Sadiek},
  \citenamefont {Alkurtass},\ and\ \citenamefont {Aldossary}}]{Sadiek2010}%
  \BibitemOpen
  \bibfield  {author} {\bibinfo {author} {\bibfnamefont {G.}~\bibnamefont
  {Sadiek}}, \bibinfo {author} {\bibfnamefont {B.}~\bibnamefont {Alkurtass}},\
  and\ \bibinfo {author} {\bibfnamefont {O.}~\bibnamefont {Aldossary}},\
  }\bibfield  {title} {\bibinfo {title} {Entanglement in a time-dependent
  coupled $\mathrm{XY}$ spin chain in an external magnetic field},\ }\href
  {https://doi.org/10.1103/PhysRevA.82.052337} {\bibfield  {journal} {\bibinfo
  {journal} {Phys. Rev. A}\ }\textbf {\bibinfo {volume} {82}},\ \bibinfo
  {pages} {052337} (\bibinfo {year} {2010})}\BibitemShut {NoStop}%
\bibitem [{\citenamefont {Barouch}\ \emph {et~al.}(1970)\citenamefont
  {Barouch}, \citenamefont {McCoy},\ and\ \citenamefont {Dresden}}]{Barouch1}%
  \BibitemOpen
  \bibfield  {author} {\bibinfo {author} {\bibfnamefont {E.}~\bibnamefont
  {Barouch}}, \bibinfo {author} {\bibfnamefont {B.~M.}\ \bibnamefont {McCoy}},\
  and\ \bibinfo {author} {\bibfnamefont {M.}~\bibnamefont {Dresden}},\
  }\bibfield  {title} {\bibinfo {title} {Statistical mechanics of the
  $\mathrm{XY}$ model. i},\ }\href {https://doi.org/10.1103/PhysRevA.2.1075}
  {\bibfield  {journal} {\bibinfo  {journal} {Phys. Rev. A}\ }\textbf {\bibinfo
  {volume} {2}},\ \bibinfo {pages} {1075} (\bibinfo {year} {1970})}\BibitemShut
  {NoStop}%
\bibitem [{\citenamefont {Barouch}\ and\ \citenamefont
  {McCoy}(1971)}]{Barouch2}%
  \BibitemOpen
  \bibfield  {author} {\bibinfo {author} {\bibfnamefont {E.}~\bibnamefont
  {Barouch}}\ and\ \bibinfo {author} {\bibfnamefont {B.~M.}\ \bibnamefont
  {McCoy}},\ }\bibfield  {title} {\bibinfo {title} {Statistical mechanics of
  the $\mathrm{XY}$ model. ii. spin-correlation functions},\ }\href
  {https://doi.org/10.1103/PhysRevA.3.786} {\bibfield  {journal} {\bibinfo
  {journal} {Phys. Rev. A}\ }\textbf {\bibinfo {volume} {3}},\ \bibinfo {pages}
  {786} (\bibinfo {year} {1971})}\BibitemShut {NoStop}%
\bibitem [{\citenamefont {Schr\"odinger}(1935)}]{Schodinger}%
  \BibitemOpen
  \bibfield  {author} {\bibinfo {author} {\bibfnamefont {E.}~\bibnamefont
  {Schr\"odinger}},\ }\bibfield  {title} {\bibinfo {title} {Discussion of
  probability relations between separated systems},\ }\href
  {https://doi.org/10.1017/S0305004100013554} {\bibfield  {journal} {\bibinfo
  {journal} {Proc. Cambridge Philos. Soc.}\ }\textbf {\bibinfo {volume} {31}},\
  \bibinfo {pages} {553} (\bibinfo {year} {1935})}\BibitemShut {NoStop}%
\bibitem [{\citenamefont {Einstein}\ \emph {et~al.}(1935)\citenamefont
  {Einstein}, \citenamefont {Podolsky},\ and\ \citenamefont {Rosen}}]{EPR}%
  \BibitemOpen
  \bibfield  {author} {\bibinfo {author} {\bibfnamefont {A.}~\bibnamefont
  {Einstein}}, \bibinfo {author} {\bibfnamefont {D.}~\bibnamefont {Podolsky}},\
  and\ \bibinfo {author} {\bibfnamefont {N.}~\bibnamefont {Rosen}},\ }\bibfield
   {title} {\bibinfo {title} {Can quantum-mechanical description of physical
  reality be considered complete?},\ }\href
  {https://doi.org/10.1103/PhysRev.47.777} {\bibfield  {journal} {\bibinfo
  {journal} {Phys. Rev.}\ }\textbf {\bibinfo {volume} {47}},\ \bibinfo {pages}
  {777} (\bibinfo {year} {1935})}\BibitemShut {NoStop}%
\bibitem [{\citenamefont {Wiseman}\ \emph {et~al.}(2007)\citenamefont
  {Wiseman}, \citenamefont {Jones},\ and\ \citenamefont
  {Doherty}}]{Wiseman2007}%
  \BibitemOpen
  \bibfield  {author} {\bibinfo {author} {\bibfnamefont {H.~M.}\ \bibnamefont
  {Wiseman}}, \bibinfo {author} {\bibfnamefont {S.~J.}\ \bibnamefont {Jones}},\
  and\ \bibinfo {author} {\bibfnamefont {A.~C.}\ \bibnamefont {Doherty}},\
  }\bibfield  {title} {\bibinfo {title} {Steering, entanglement, nonlocality,
  and the einstein-podolsky-rosen paradox},\ }\href
  {https://doi.org/10.1103/PhysRevLett.98.140402} {\bibfield  {journal}
  {\bibinfo  {journal} {Phys. Rev. Lett.}\ }\textbf {\bibinfo {volume} {98}},\
  \bibinfo {pages} {140402} (\bibinfo {year} {2007})}\BibitemShut {NoStop}%
\bibitem [{\citenamefont {Costa}\ and\ \citenamefont
  {Angelo}(2016)}]{Costa2016}%
  \BibitemOpen
  \bibfield  {author} {\bibinfo {author} {\bibfnamefont {A.~C.~S.}\
  \bibnamefont {Costa}}\ and\ \bibinfo {author} {\bibfnamefont {R.~M.}\
  \bibnamefont {Angelo}},\ }\bibfield  {title} {\bibinfo {title}
  {Quantification of einstein-podolsky-rosen steering for two-qubit states},\
  }\href {https://doi.org/10.1103/PhysRevA.93.020103} {\bibfield  {journal}
  {\bibinfo  {journal} {Phys. Rev. A}\ }\textbf {\bibinfo {volume} {93}},\
  \bibinfo {pages} {020103} (\bibinfo {year} {2016})}\BibitemShut {NoStop}%
\bibitem [{\citenamefont {Wang}\ \emph {et~al.}(2017)\citenamefont {Wang},
  \citenamefont {Tang}, \citenamefont {Wei}, \citenamefont {Yu}, \citenamefont
  {Ke}, \citenamefont {Xu}, \citenamefont {Li},\ and\ \citenamefont
  {Guo}}]{SQC-Lab1}%
  \BibitemOpen
  \bibfield  {author} {\bibinfo {author} {\bibfnamefont {Y.-T.}\ \bibnamefont
  {Wang}}, \bibinfo {author} {\bibfnamefont {J.-S.}\ \bibnamefont {Tang}},
  \bibinfo {author} {\bibfnamefont {Z.-Y.}\ \bibnamefont {Wei}}, \bibinfo
  {author} {\bibfnamefont {S.}~\bibnamefont {Yu}}, \bibinfo {author}
  {\bibfnamefont {Z.-J.}\ \bibnamefont {Ke}}, \bibinfo {author} {\bibfnamefont
  {X.-Y.}\ \bibnamefont {Xu}}, \bibinfo {author} {\bibfnamefont {C.-F.}\
  \bibnamefont {Li}},\ and\ \bibinfo {author} {\bibfnamefont {G.-C.}\
  \bibnamefont {Guo}},\ }\bibfield  {title} {\bibinfo {title} {Directly
  measuring the degree of quantum coherence using interference fringes},\
  }\href {https://doi.org/10.1103/PhysRevLett.118.020403} {\bibfield  {journal}
  {\bibinfo  {journal} {Phys. Rev. Lett.}\ }\textbf {\bibinfo {volume} {118}},\
  \bibinfo {pages} {020403} (\bibinfo {year} {2017})}\BibitemShut {NoStop}%
\bibitem [{\citenamefont {Zhang}\ \emph {et~al.}(2018)\citenamefont {Zhang},
  \citenamefont {Liu}, \citenamefont {Yu},\ and\ \citenamefont
  {Tong}}]{SQC-Lab2}%
  \BibitemOpen
  \bibfield  {author} {\bibinfo {author} {\bibfnamefont {D.-J.}\ \bibnamefont
  {Zhang}}, \bibinfo {author} {\bibfnamefont {C.~L.}\ \bibnamefont {Liu}},
  \bibinfo {author} {\bibfnamefont {X.-D.}\ \bibnamefont {Yu}},\ and\ \bibinfo
  {author} {\bibfnamefont {D.~M.}\ \bibnamefont {Tong}},\ }\bibfield  {title}
  {\bibinfo {title} {Estimating coherence measures from limited experimental
  data available},\ }\href {https://doi.org/10.1103/PhysRevLett.120.170501}
  {\bibfield  {journal} {\bibinfo  {journal} {Phys. Rev. Lett.}\ }\textbf
  {\bibinfo {volume} {120}},\ \bibinfo {pages} {170501} (\bibinfo {year}
  {2018})}\BibitemShut {NoStop}%
\bibitem [{\citenamefont {Yu}\ and\ \citenamefont {G\"uhne}(2019)}]{SQC-Lab3}%
  \BibitemOpen
  \bibfield  {author} {\bibinfo {author} {\bibfnamefont {X.-D.}\ \bibnamefont
  {Yu}}\ and\ \bibinfo {author} {\bibfnamefont {O.}~\bibnamefont {G\"uhne}},\
  }\bibfield  {title} {\bibinfo {title} {Detecting coherence via spectrum
  estimation},\ }\href {https://doi.org/10.1103/PhysRevA.99.062310} {\bibfield
  {journal} {\bibinfo  {journal} {Phys. Rev. A}\ }\textbf {\bibinfo {volume}
  {99}},\ \bibinfo {pages} {062310} (\bibinfo {year} {2019})}\BibitemShut
  {NoStop}%
\bibitem [{\citenamefont {Singha~Roy}\ \emph {et~al.}(2020)\citenamefont
  {Singha~Roy}, \citenamefont {Santalla}, \citenamefont
  {Rodr\'{\i}guez-Laguna},\ and\ \citenamefont {Sierra}}]{Roy2020}%
  \BibitemOpen
  \bibfield  {author} {\bibinfo {author} {\bibfnamefont {S.}~\bibnamefont
  {Singha~Roy}}, \bibinfo {author} {\bibfnamefont {S.~N.}\ \bibnamefont
  {Santalla}}, \bibinfo {author} {\bibfnamefont {J.}~\bibnamefont
  {Rodr\'{\i}guez-Laguna}},\ and\ \bibinfo {author} {\bibfnamefont
  {G.}~\bibnamefont {Sierra}},\ }\bibfield  {title} {\bibinfo {title}
  {Entanglement as geometry and flow},\ }\href
  {https://doi.org/10.1103/PhysRevB.101.195134} {\bibfield  {journal} {\bibinfo
   {journal} {Phys. Rev. B}\ }\textbf {\bibinfo {volume} {101}},\ \bibinfo
  {pages} {195134} (\bibinfo {year} {2020})}\BibitemShut {NoStop}%
\end{thebibliography}%
\end{document}